\documentclass[letterpaper,twocolumn,10pt]{article}
\usepackage{usenix2019_v3}

\usepackage{cite}
\usepackage{subcaption}
\usepackage{csquotes}
\usepackage{tabularx}
\usepackage{booktabs}
\usepackage{comment}
\usepackage{multirow}
\usepackage{amsmath}
\usepackage{soul}
\usepackage{xspace}
\usepackage{xcolor,colortbl}
\usepackage{xcolor}
\usepackage{soul}
\usepackage{stfloats}
\usepackage{bbm}

\definecolor{Gray}{gray}{0.85}
\definecolor{LightCyan}{rgb}{0.88,1,1}

\newcolumntype{a}{>{\columncolor{Gray}}r}
\newcolumntype{m}{>{\hsize=0.70\hsize}X}
\newcolumntype{v}{>{\hsize=0.20\hsize}X}
\newcolumntype{b}{X}
\newcolumntype{s}{>{\hsize=0.30\hsize}X}

\usepackage{graphicx}
\usepackage{array}
\usepackage{multirow}
\usepackage{pifont}
\usepackage{color, colortbl}
\usepackage{siunitx}
\usepackage{booktabs}
\usepackage{rotating}
\usepackage{xparse}
\usepackage{tabularx}
\usepackage{url}
\usepackage{pifont}
\usepackage[font=small]{caption}
\usepackage{subcaption}
\usepackage{tikz}
\usepackage{makecell}
\usepackage{xspace}

\newcommand*{\escape}[1]{\texttt{\textbackslash#1}}
\newcommand*{\escapeI}[1]{\texttt{\expandafter\string\csname #1\endcsname}}

\begin{document}
%
\title{Tracking the Takes and Trajectories of English-Language News Narratives across Trustworthy and Worrisome Websites}

\thispagestyle{plain}
\pagestyle{plain}
\author{
{\rm Hans W. A. Hanley }\\
Stanford University 
\and
{\rm Emily Okabe}\\
Stanford University 
 \and
 {\rm Zakir Durumeric}\\
Stanford University
} 

\maketitle

\begin{abstract}
Understanding how misleading and outright false information enters news ecosystems remains a difficult challenge that requires tracking how narratives spread across thousands of fringe and mainstream news websites. To do this, we introduce a system that utilizes encoder-based large language models and zero-shot stance detection to scalably identify and track news narratives and their attitudes across over 4,000~factually unreliable, mixed-reliability, and factually reliable English-language news websites. Running our system over an 18~month period, we track the spread of 146K~news stories. Using network-based interference via the NETINF algorithm, we show that the paths of news narratives and the stances of websites toward particular entities can be used to uncover slanted propaganda networks (\textit{e.g.}, anti-vaccine and anti-Ukraine) and to identify the most influential websites in spreading these attitudes in the broader news ecosystem. We hope that increased visibility into our distributed news ecosystem can help with the reporting and fact-checking of propaganda and disinformation.
\end{abstract}

\section{Introduction}

Misinformation has promoted dangerous fake health cures~\cite{ball2020epic}, promoted jingoism and propaganda during wars~\cite{Leonhardt2023,pierri2023propaganda,shahi2024warclaim}, and incited violence~\cite{banaji2019whatsapp,aliapoulios2022gospel}. While there has been significant investigation into how misleading information spreads across social media platforms and fringe websites~\cite{kwon2013prominent,kwon2013aspects,starbird2018ecosystem}, recent work has emphasized that the vast majority of people do not visit fringe websites or regularly encounter misinformation on social media~\cite{moore2023exposure, allen2020evaluating}. Rather, most people consume news through more mainstream platforms like television news~\cite{allen2020evaluating}. However, systematically tracking how misleading, propagandistic, and outright false information spreads from untrustworthy websites into mainstream media and how fringe sites influence the broader news ecosystem remains a significant technical challenge due to the magnitude and distributed nature of the global news ecosystem~\cite{biancovilli2021misinformation,hanley2022happenstance, starbird2018ecosystem, allcott2017social}. 


In this work, we introduce and validate a system for scalably identifying and tracking potentially unreliable news narratives across different English-language media ecosystems. Building on past work~\cite{hanley2023specious, mccoy2012pharmaleaks, afroz2014doppelganger, zeng2020bad}, our proposed approach: (1)~collects articles by continually crawling news sites from across media ecosystems; (2)~extracts semantic {narratives} and sites' stances towards different topics using a 
fine-tuned version of the {e5-base-v2} large language model~\cite{wang2022text}, DP-Means clustering~\cite{dinari2022revisiting}, and zero-shot stance detection~\cite{allaway2020zero}; and (3)~identifies the relationships between news sites and broader ecosystems using the NETINF algorithm~\cite{gomez2012inferring}.
We emphasize that our approach does not make factual assessments of individual stories, which is a deeply nuanced task. Rather, our system allows us to shed light on how stories travel across the distributed news ecosystem.\looseness=-1

We analyze the results from our deployed system for an 18~month period during which we collected articles from pre-curated lists of 1,003~ factually unreliable news websites (\textit{e.g.}, twisted.news),  1,012~mixed factuality reliability websites (\textit{e.g.}, foxnews.com), and 2,061~factually reliable news websites (\textit{e.g.}, washingtonpost.com) maintained by Media-Bias/Fact-Check~\cite{mediabias2023}.
Analyzing 146K~stories that our system extracted from 29M~articles on these news websites, we observe significant crossover in the stories covered by different news ecosystems~\cite{vosoughi2018spread}. We show that reliable and mixed-reliability news websites play the largest role in setting the stories and narratives addressed by other websites.  However, despite covering similar topics, our stance analysis reveals that each type of website adopts distinctive stances towards shared topics, with reliable news websites generally being left-leaning and pro-Ukraine and unreliable websites being the most right-leaning and anti-Ukraine.


Framing our story clusters as cascades, our system uses NETINF~\cite{gomez2012inferring} to uncover relationships between news sites and to detect potential networks of coordinating websites that spread particular slanted content and narratives. For example, using this approach we identify a network of right-leaning news websites that ostensibly act as local-news websites all operated by Metric Media, LLC\@. From the outputted results, we further identify the sites most influential in spreading stories to unreliable sites (\textit{e.g.}, thegatewaypundit.com) and the sites from which both reliable and unreliable news websites most commonly adopt stories (\textit{e.g.}, dailymail.co.uk and ussanews.com). 
Additionally, we identify the sites that most effectively promote specific types of information across ecosystems like anti-vaccine misinformation (naturalnews.com, theepochtimes.com, and vaccines.news) and anti-Ukrainian propaganda (rt.com, sputniknews.com, and news-front.info). 


Ultimately, our work introduces an end-to-end system for building a near-global perspective of the English-language news ecosystem and explores how tracking how narratives travel within it can help us to understand how misleading information enters mainstream news and to uncover previously unknown relationships between news sites. We hope that our approach can serve as the foundation for further study of how information spreads online. Our code and URL data is available at \url{https://github.com/hanshanley/tracking-takes}.

\section{Related Work}

Significant prior work has studied news ecosystems and analyzed how misinformation spreads online. Here, we summarize the prior work that our study builds on:

\vspace{2pt}
\noindent
\textbf{Tracking Narratives on News Websites.}\quad Several studies have utilized online document clustering~\cite{yin2018model,blei2010nested} for tracking news stories. For example, Zhang et~al.~\cite{zhang2022unsupervised} identify potential events by monitoring for the appearance of specific phrases or keywords, cluster identified phrases that may indicate news events, and train a series of classifiers to assign news articles to identified clusters. Similarly, by clustering a collection of short phrases or ``memes,'' Leskovec et~al.\ find that smaller blogs often play a definitive role in encouraging the adoption of particular language onto mainstream websites~\cite{leskovec2009meme}. Rodriguez et~al.~\cite{gomez2013structure,gomez2012inferring} examine the changing relationships between websites during the discussion of news events, finding that connections between websites increase during periods of high activity.

While many studies have analyzed topics using statistical word-association approaches like Latent Dirichlet Allocation (LDA) and Dynamic Topic Models~\cite{albalawi2020using,momeni2018modeling,zhou-etal-2015-unsupervised}, recent works such as those by Meng et~al.~\cite{meng2022topic}, Hanley et~al.~\cite{hanley2022happenstance, hanley2024machine}, and Grootendorst~\cite{grootendorst2022bertopic} have used large language models (LLMs) for more granular topic modeling. In line with our work, Nakshatri et~al.~\cite{nakshatri2023using} utilize peak detection and HDBSCAN~\cite{mcinnes2017hdbscan} on news article embeddings to identify the most prominent news events in a stream of news articles. Saravanakumar et al.~\cite{saravanakumar2021event} similarly utilize an external named entity recognition system to embed entity knowledge into a BERT language model to differentiate between news articles about different events.  Beyond these quantitative approaches, many prior works have qualitatively investigated the spread of individual news stories (\textit{e.g.},~\cite{schafer2021climate,starbird2018ecosystem,prochaska2023mobilizing}).

Most similar to our work, Hanley et al.~\cite{hanley2023specious}, using MPNet and DP-Means clustering, track news narratives across a smaller number of fringe websites to determine the role that individual unreliable news websites play in originating and amplifying news narratives. Their work finds that less-popular websites oftentimes play an outsized role in promoting narratives that reverberate across the unreliable news ecosystem. 

In contrast to these prior works, our study accounts for the \emph{stance} towards each topic in order to better differentiate between articles that cover the same topic. Tracking stance enables our work to understand the widespread understanding of individual websites' ideologically skew, changes in coverage of individual topics, and the detection of websites that coordinate in spreading particular types of propaganda.


\vspace{2pt}
\noindent
\textbf{Analyzing the Spread of Misinformation.}\quad
While our approach is one of the first to track both topic and valence/stance towards that topic in a programmatic manner, several prior works have focused on the peculiarities, detection, and the spread of misinformation. For example, Ma et~al.~\cite{ma2016detecting} and Jin et~al.~\cite{jin2017multimodal} utilize recurrent neural networks to analyze and detect the spread of unreliable rumors on social media. Abdali~\cite{abdali2021identifying} et~al., taking a domain-based approach, use website screenshots to assess the credibility of news websites. In addition to analyzing the spread of general misinformation on particular social platforms, other works have further investigated the spread of specific narratives, including those concerning the Syrian White Helmets~\cite{starbird2018ecosystem}, QAnon~\cite{amarasingam2020qanon,hanley2022no,papasavva2021qoincidence}, the Russo-Ukrainian War~\cite{hanley2022happenstance,hanley2022special,pierri2023propaganda}, and COVID-19~\cite{madraki2021characterizing,cuan2020misinformation,aghababaeian2020alcohol}. We note that because work utilizes topic analysis followed by stance detection, our system can be used to quickly identify websites and topics that deserve in-depth investigation of particular types of coverage of individual events.

Building off these studies, several works have analyzed the characteristics of misinformation. Juul and Ugander find that often false information on Twitter spreads faster and wider than factual information~\cite{juul2021comparing}. Indeed, Kwon et~al.~\cite{kwon2013prominent}, utilizing the distinct temporal differences between reliable information and unreliable rumors, are able to classify these rumors with an $F_1$-score as high as 0.878. In a different work~\cite{kwon2013aspects}, Kwon et~al.\ analyze the semantic and structural characteristics of rumors on Twitter. In a different vein, Using a learning-to-rank-based approach and ClaimBuster API, Paudel et al.~\cite{paudel2023lambretta} identify potential claims that should be fact-checked on Twitter~\cite{hassan2017claimbuster}. Beyond studying the dynamics of misinformation, Bak et~al.~\cite{bak2022combining} have proposed concrete steps to ameliorate the spread of misinformation, including removal and nudges. Finally, Kaiser et~al.~\cite{kaiser2021adapting} have studied how borrowing techniques from the security warning landscape might help inform users of potential misinformation.

Unlike the past approaches outlined above, by utilizing fine-tuned encoder-based large language models, our work scalably tracks and identifies unique news stories across thousands of news websites without depending on particular keywords or by limiting analysis to a subset of unreliable websites previously fact-checked or curated by experts~\cite{starbird2018ecosystem,hanley2023specious}. By utilizing network analysis combined with stance detection, our work further provides a highly interpretable means of understanding the spread and dynamics of propagandistic, biased, or factually unreliable narratives across multiple types of media ecosystems.

\section{Methodology}

In this section, we provide an overview of our data collection and approach for extracting and tracking narratives across different types of news websites. 

\subsection{News Websites}

Our study analyzes articles collected from three sets of English-language news websites of varying factual reliability. We specifically track narratives on websites rated by Media-Bias/Fact-Check~\cite{mediabias2023}, a media monitoring website founded by Dave M. Van Zandt to assess the factual reliability of individual websites given its widespread use in prior work~\cite{weld2021political,hanley2023specious,babaei2021analyzing,nakov2021fake} and its ratings' high agreement with other organizations like NewsGuard.

\begin{figure*}
  \centering
  \includegraphics[width=1.0\linewidth]{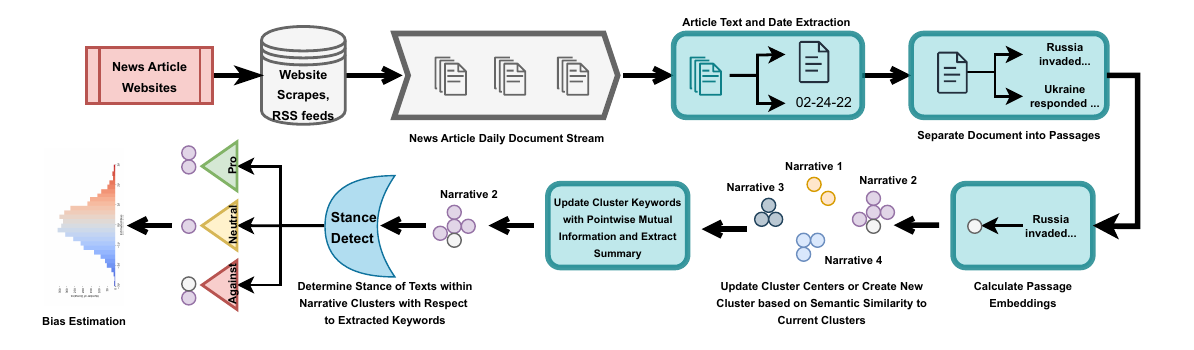}
\vspace{-15pt}
\caption{Our pipeline for identifying, labeling, and extracting the stance of story clusters from the daily publications of news websites.}
\label{fig:pipeline}
\end{figure*}

\vspace{2pt}
\noindent
\textbf{Unreliable News Websites.}\quad
We collect news articles from 1,003~websites labeled as having ``low'' or ``very low'' factual reporting by Media-Bias/Fact-Check~\cite{mediabias2023}. We extend this list with conspiracy theory-promoting websites identified by Hanley et al.~\cite{hanley2022golden}. Our list of \textit{unreliable} news websites includes pseudo-science sites like vaccine.news, state-propaganda outlets such as rt.com, and partisan websites with low-factuality ratings like the liberal-leaning occupydemocrats.com.

\vspace{2pt}
\noindent
\textbf{Mixed-Reliability News Websites.}\quad
We collect articles from 1,012~\textit{mixed-reliability} news websites labeled as having ``mixed'' factual reporting by Media-Bias/Fact-Check~\cite{mediabias2023}. This list includes websites across the political spectrum, such as foxnews.com, nypost.com, and theguardian.com.
 
\vspace{2pt}
\noindent
\textbf{Reliable News Websites.}\quad We collect articles from 2,061~\textit{reliable} news websites labeled as having ``high'', ``very high'', or ``mostly factual'' reporting by Media-Bias/Fact-Check~\cite{mediabias2023}. The category ``mostly factual'' is included to capture sources with strong reputations like The Washington Post. This list features websites such as reuters.com and apnews.com.

\vspace{3pt}
\noindent
We lastly note that we utilize the full set of English-language news websites from the lists of Media-Bias/Fact-Check~\cite{mediabias2023} and Hanley et al.~\cite{hanley2022golden} that were accessible to us from the beginning of our study.
\subsection{Definition of a News Story}

Our approach tracks specific news stories and their propagation across websites rather than analyzing broader themes as captured by methods like LDA~\cite{allan2002detection,devine2022unsupervised,jelodar2019latent}.
Following previous research~\cite{hanley2023specious,hanley2022happenstance}, we adopt Event Registry’s definition of a \emph{news story} as ``collections of documents that seek to address the same \textit{event} or \textit{issue}''~\cite{leban2014event,miranda-etal-2018-multilingual}. It is important to note that even if two ideas are related, they may not constitute the same news story. For example, while ``Florida Governor Ron DeSantis declares for President'' and ``Nikki Haley surpasses Ron DeSantis in the polls'' are related, they are considered separate news stories in our work.

\subsection{System Architecture}

Our approach for capturing and tracking news stories builds on the LLM-based narrative tracking methodology introduced by Hanley et~al.~\cite{hanley2023specious}. However, while Hanley et~al.'s method scalably tracks individual topics, their work does not incorporate articles' attitudes towards a topic. While this was not problematic for their work, which focused on the spread of stories amongst \textit{unreliable} news websites, the approach cannot track news stories across a broader set of news websites that present stories in dramatically different ways. We expand their method to additionally account for the \emph{stance}/valence of news articles towards a topic (\textit{i.e.}, we distinguish between articles that cover vaccines positively vs.\ negatively). 

As shown in Figure~\ref{fig:pipeline}, our system identifies stories by: (1)~scraping articles from news sites, (2)~splitting articles into passages of 100~words~\cite{piktus2021web,hanley2023specious}, (3)~embedding passages with a fine-tuned LLM~\cite{wang2022text}, and (4)~clustering news articles using an optimized version of the DP-Means algorithm~\cite{johnson2019billion,dinari2022revisiting}. To describe clusters that each represent a story, we extract keywords from the resulting cluster using Pointwise Mutual Information (PMI) and performing multi-document summarization with an open source LLM\@. 
Building on the clusters, we utilize network inference techniques to identify website relationships and \textit{zero-shot} stance detection~\cite{allaway2020zero,hanley2023tata} to determine the stance/position of individual passages within each cluster. Finally, based on individual websites' stances toward their given topics, we perform bias estimation to quantify websites' biases along various political and non-political axes.  
We detail each stage below:

\vspace{2pt}
\noindent
\textbf{Collecting and Preparing News Articles.}\quad \label{sec:preprocess}
We crawl our set of 4,076~websites daily using the Go Colly library~\cite{smith2019go} from January 1, 2022 to July 1, 2023. Each day, we collect every site's homepage, RSS feeds, and linked articles. We collected a total 29.0M~articles: 17.9M~articles from reliable news sites (median 2,467~articles/site), 8.7M~articles from mixed-reliability news sites (median 964~articles/site), and 2.5M~articles from unreliable news websites (median 219~articles/site). We will provide to URLs researchers on request.\looseness=-1
 

\begin{table}
    \centering
    \small
    \begin{tabular}{cccccc}
    \toprule
  {}  & {}&{all-mpnet} &{all-mpnet}&{e5-base-v2} \\
  {BERT}  & {USE}&{specious} &{peft+lora}&{peft+lora} \\
  \midrule
  0.464 & 0.749 &0.856  &0.860 &\textbf{0.866}\\
  \bottomrule
    \end{tabular}
    \vspace{-5pt}
    \caption{Model Performance on SemEval STS Benchmark. Our PEFT+LoRA models fine-tuned using unsupervised contrastive loss perform better than prior work~\cite{cer2017semeval,devlin2019bert,cer2018universal,reimers2019sentence,hanley2023specious}.}
    \label{tab:sts-benchmark}
\end{table}

To prepare our news article data for embedding, we first remove any URLs, emojis, and HTML tags from the text. Then, inline with prior work, after first separating articles into paragraphs by splitting text on ({\escape{n}}) or tab ({\escape{t}}) characters~\cite{hanley2022partial}, we subsequently divide paragraph into constituent \textit{passages} with at most 100~words~\cite{piktus2021web,hanley2022happenstance,hanley2022partial}. This enables us to fit passages into the context window of our LLM embedding model. Further, given that articles often address multiple ideas, embedding {passages} allows us to track the often single idea present within the passage~\cite{piktus2021web,hanley2022partial}. Our dataset consists of 428M~passages. For additional details, see Appendix~\ref{sec:appendix-pre-proces}.

\vspace{2pt}
\noindent
\textbf{Embedding Passages.}\quad 
Before embedding our articles' passages, to ensure that our embedding model is attuned to the language of news articles, we tailor our model to our domain of our collected articles using \textit{Parameter Efficient Fine-Tuning}/PEFT~\cite{lester2021power} through \textit{Low-Rank Adaption}/LoRA~\cite{hu2021lora} with an unsupervised contrastive learning step based on SimCSE~\cite{gao2021simcse}. Rather than directly fine-tuning the original model's weights as in Hanley et~al.~\cite{hanley2023specious}, this approach freezes the originally trained large language model and introduces an additional set of parameters of reduced dimensionality that are then fine-tuned for purposes, allowing for better generalizability~\cite{hu2021lora}.  We utilize default LoRA hyperparameters of rank=8 and $\alpha$=16.\footnote{\url{https://huggingface.co/docs/peft/task_guides/semantic-similarity-lora}}  See Appendix~\ref{sec:peft} and~\ref{sec:contrastive} for additional details. We utilize cosine similarity of embeddings to determine passages' estimated semantic similarity~\cite{chandrasekaran2021evolution,gao2021simcse,piktus2021web,grootendorst2022bertopic}.

We specifically fine-tune and evaluate two public open-source large language models, \texttt{e5-base-v2}~\cite{wang2022text} and \texttt{MPNet}~\cite{song2020mpnet} using this approach. We benchmark these two fine-tuned models on the SemEval STS-benchmark (Table~\ref{tab:sts-benchmark}) and find that the approach outperforms prior work as general models for semantic similarity. We use the fine-tuned \texttt{e5-base-v2} model in this work given its top performance. 

\vspace{2pt}
\noindent
\textbf{Story Identification.\label{sec:idstories}}\quad We base our story-identification algorithm on Dinari et~al.'s optimized and parallelizable version of the DP-Means algorithm, a non-parametric version of K-means~\cite{dinari2022revisiting}. We utilize this approach as it is highly scalable (able to cluster our 428M embeddings) unlike other LLM-based approaches~\cite{grootendorst2022bertopic} while also allowing us to identify stories without 
\textit{a priori} knowledge. To further scale the approach, we re-implement DP-Means~\cite{dinari2022revisiting} to use the GPU-enhanced {FAISS} library~\cite{johnson2019billion} to perform the embedding-to-cluster assignments and similarity calculations required by DP-Means. To determine a suitable threshold for clustering two news passages together, after fine-tuning \texttt{e5-base-v2}, we benchmark our model on the English portion SemEval 2022 Task~8 dataset~\cite{goel2022semeval}. The SemEval 2022 Task~8 dataset consists of two parallel lists of news articles where each pair is graded on whether they are about the same news story. Our model achieves a max $F_1$-score of 0.793 on this dataset near a cosine similarity threshold of 0.50, which we use in this work. We provide examples of passage pairs in Appendix~\ref{sec:thresholds}.

From January~1, 2022 to July~1, 2023, clustering all our embeddings required the equivalent of 12~days using an NVIDIA A100 GPU\@. After clustering, like in other works~\cite{leskovec2009meme,hanley2023specious}, we filter out clusters where 50\% or more of the passages are from only one website (\textit{e.g.}, website-specific headers or author bios). After this pruning, we identified 146,212 story clusters. We provide 30~cluster examples in Appendix~\ref{sec:cluster-eval} and evaluate these 30~clusters to ensure that they contain coherent stories using the method outlined by Hanley et~al.~\cite{hanley2023specious}. We achieve an estimated precision of 99.3\% of assigning passages to appropriate story clusters where each passage matches the summary, keywords, and other passages in the cluster.

\vspace{2pt}
\noindent
\textbf{Story Summarization and Labeling.}\quad
To build human-understandable representations of our clusters, we extract keywords using pointwise mutual information (PMI), an information-theoretic for uncovering associations~\cite{bouma2009normalized}, to uncover the words most associated with each story cluster~\cite{hanley2022special}. To make these words more uniform, we lemmatize each word to each cluster before calculating PMI\@. For details on PMI, see Appendix~\ref{sec:appendix-pmi}.  In addition, we perform multi-document summarization utilizing an instruction fine-tuned version of Llama 3~\cite{dubey2024llama}.\footnote{\url{https://huggingface.co/meta-llama/Meta-Llama-3.1-8B-Instruct}} This enables us to summarize the different perspectives of the passages within a given cluster, while also allowing humans to easily understand a story cluster's contents. We utilize the following prompt to summarize the contents of each of our clusters: \textit{You work for a news researcher and your job is to summarize articles. Write a single concise collective abstractive summary of the texts, where individual texts are separated by |||||, and return your response as a single summary that covers the key points of the text.}

\vspace{2pt}
\noindent
\textbf{Website Relationship Inference.\label{sec:back-web-relationships}}\quad
To further understand the relationships between news sites, we analyze how stories spread across websites over time. We consider the set of articles in a cluster as a time cascade based on the date that each article was published, and we use an open source version of {NETINF}~\cite{gomez2012inferring} to infer the underlying structure and relationship amongst our set of news websites.\footnote{\url{https://snap.stanford.edu/netinf/}} Given a set of time cascades (\textit{e.g.,} the time steps for when a particular website posts an article within a given story cluster), while assuming that each node in a particular cascade is influenced by exactly one other node, the {NETINF} algorithm attempts to infer the optimal network to explain the observed posting behavior~\cite{gomez2012inferring}. Based on each website posting behavior across the different cascades, {NETINF} estimates the number of times that each website copied information from another as well as the time delay between copies. 
We provide additional details in Appendix~\ref{sec:netinf}.

\vspace{2pt}
\noindent
\textbf{Stance Detection.\label{sec:stance-detection}}\quad
While passages may cover the same story, they often adopt different \textit{stances}~\cite{mohammad2016semeval, kuccuk2020stance, hanley2022happenstance} in addressing the same event. After identifying the stories on our set of news websites, we employ stance detection to understand how different websites address each story. Stance detection methods determine the attitude of an author toward a specific topic or target~\cite{biber1988adverbial}. Typically, stance detection involves taking a passage $p_i$ and a topic or target $t_i$, and outputting the stance $s_i \in \{Pro, Against, Neutral\}$ of the passage towards the target, where the target is a \textit{noun} or a \textit{noun phrase}. Given that most stance detection methods heavily rely on the topic or target, with many models struggling to generalize to topics or targets outside their domain, various models have been developed to perform stance detection in \textit{zero-shot} (where the tested topics or targets are not in the training data) and \textit{few-shot} (where very few examples of the tested topics or targets are in the training data) settings~\cite{allaway2020zero, liang2022jointcl}.

\begin{figure}
  \centering
  \vspace{-5pt}
  \includegraphics[width=\columnwidth]{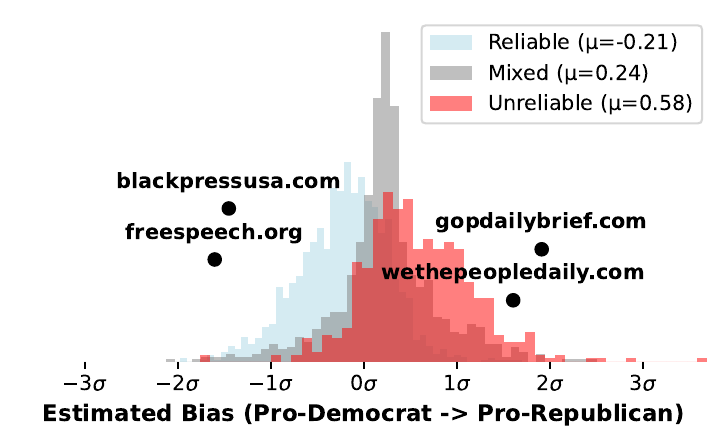}
  \vspace{-20pt}
\caption{Estimated Partisanship via Bayesian regression of our websites based on their stances to articles' topics.}
\label{fig:estimated-partisanship}
\end{figure}

To perform this stance detection, we utilize the current state-of-the-art zero-shot {TATA} model~\cite{hanley2023tata}, which was trained on the {VAST} dataset~\cite{allaway2020zero}. We note that the size of our dataset of stance pairs precluded us from using popular large language model services like GPT-4 or Claude Sonnet. To enhance this model, we retrained it on both the {VAST} dataset and  news-specific stance detection NewsMTSC dataset~\cite{Hamborg2021b}. By training the {TATA} model using this extended dataset, we achieved state-of-the-art $F_1$ scores of $0.781$ in the zero-shot setting and $0.741$ in the few-shot setting on the {VAST} test dataset, and a macro $F_1$ score of $0.849$ on the {NewsMTSC} test dataset.

Specifically, rather than performing stance detection on a pre-determined set of topics~\cite{grimminger2021hate,lai2018stance,kuccuk2020stance}, we leverage our topic and story modeling to conduct stance detection across each story cluster. Once we extract story keywords using PMI, we utilize the Python \texttt{NLTK} library's Part-of-Speech (POS) tagging function to identify the most distinctive noun keywords~\cite{bird2009natural}, capturing the topic addressed in each passage. We further use the \texttt{NLTK} library to filter out common first names (\textit{e.g.}, Michael, Jessica) from our stance detection algorithm and employ the Python \texttt{spaCy} library~\cite{vasiliev2020natural} to exclude nouns that fall into the following categories: \textit{FAC, LOC, WORK\_OF\_ART, DATE, TIME, PERCENT, MONEY, QUANTITY, ORDINAL, CARDINAL}. This approach ensures that passages are not erroneously categorized as \textit{Pro} or \textit{Against} particular dates or monetary amounts. To ensure robust measurements of the collective ecosystems' and websites' stances toward specific entities, we gather the top 5,000 noun entities from our data and perform stance detection on each passage within each cluster where it appears among the top 10 PMI keywords. Altogether, this process involves running stance detection on 96.3M~passage and keyword pairs, requiring the equivalent of 15~days of computation on a NVIDIA A100.




\vspace{2pt}
\noindent
\textbf{Interpretable Mapping of Websites' Biases.\label{sec:bias-mapping}}\quad
A simplistic approach to understanding a website's overall bias (\textit{i.e} how anti or pro) toward an entity such as ``Ukraine'' would involve aggregating the percentage of their articles that had pro-``Ukraine'' and anti-``Ukraine'' stances (\textit{i.e.}, \% pro-Ukraine articles $-$ \%anti-Ukraine articles). However, this approach could potentially fail given that some websites may not have an abundance of articles focused on Ukraine or may only discuss Ukraine-related entities to obfuscate their bias. As such, taking inspiration from Waller et~al.~\cite{waller2021quantifying} who train Word2Vec models to predict subreddit's bias, we instead take a holistic approach by aggregating each website's respective stances to their written-about entities and predicting bias via Bayesian regression models.

\begin{table}
\centering
\small
\begin{tabular}{lll}
\toprule
 {Reliable News} &  Mixed  News &Unreliable News\\ \midrule
Pro CDC   &Against Kardashian & Against Pfizer \\

Pro Quantum  & Pro Gunnar  & Against Vaccine  \\
Pro Senate & Pro Alnassar & Against Wuhan \\
\bottomrule
\end{tabular}
\vspace{-5pt}
\caption{\label{tab:associated-keywords} Keywords most associated with each news ecosystem estimated using PMI.} %
\end{table}
 
To estimate websites' biases toward a subject along a given axis, we first gather a seed set of websites with at least 250~articles\footnote{This ensures that the margin of error for probabilities is below 0.10 with a 95\% confidence interval based on the normal distribution.} discussing the entity and compute their simplistic bias score (\textit{i.e.}, 
\% pro-entity articles $-$ \% anti-entity articles).  To make these values more interpretable, we normalize these scores as z-scores (\textit{i.e.}, mean 0 and variance 1), such that a score of 1.0 can be interpreted as bias in favor of entity one standard deviation above the mean~\cite{waller2021quantifying}. Following this calculation, we subsequently train a linear Bayesian regression model with $L_2$ regularization to predict this bias score by utilizing our seed set of websites' stances to other entities (besides the one in question). Finally, once trained, using the model, we estimate the rest of our websites' bias scores to the given entity. We adopt a Bayesian model approach as this directly enables us to quantify how individual stances contribute to our prediction of a given website's bias.

To validate this approach, we mapped our websites to partisanship scores along the US left--right political spectrum (Figure~\ref{fig:estimated-partisanship}) using the keywords ``democrat'' and ``republican,'' and a seed set of 105~websites.  The partisanship scores from the resulting model had a $\rho=0.51$ Spearman correlation with the partisanship labels (Far-Right, Right, Right-Center, Center, Left-Center, etc.) provided by Media-Bias/Fact-Check. As seen in Table~\ref{tab:most-partisans-keywords}, some of the most right-leaning partisan keywords included positive stances towards Dinesh D'Souza, a right-leaning commentator~\cite{Wilkie2024}, and America, while having a negative stance toward communism. On the Democratic side, the associated stances include being against Texas, conservatives, and the former Republican Congressman George Santos. Similarly, as seen in Figure~\ref{tab:most-partisans-keywords} and matching the partisan labels from Media-Bias/Fact-Check, we broadly observe that our set of reliable websites is left-leaning and the unreliable websites are right-leaning.

\begin{table}[t]

\centering
\small
\setlength\tabcolsep{4pt}
\begin{tabular}{lrl}
\toprule
{Republican Stances} & {Coeff.} & {Std.} \\
\midrule
Pro Souza & 0.311 & 0.083 \\ 
Pro America & 0.245 & 0.122 \\ 
Against Communist & 0.215 & 0.102 \\ 
\midrule
{Democratic Stances} & {Coeff.} & {Std.} \\ \midrule
Against Santos & -0.323 & 0.105 \\
Against Texas & -0.315 & 0.075 \\
Against Conservative & -0.282 & 0.098 \\
\bottomrule
\end{tabular}
\vspace{-5pt}
\caption{\label{tab:most-partisans-keywords} The stances most associated with U.S. partisanship factions, estimated using Bayseian regression.}
\end{table}


\begin{figure*}[h]
 \centering
        \centering
        \vspace{-25pt}
        \includegraphics[width=1\textwidth]{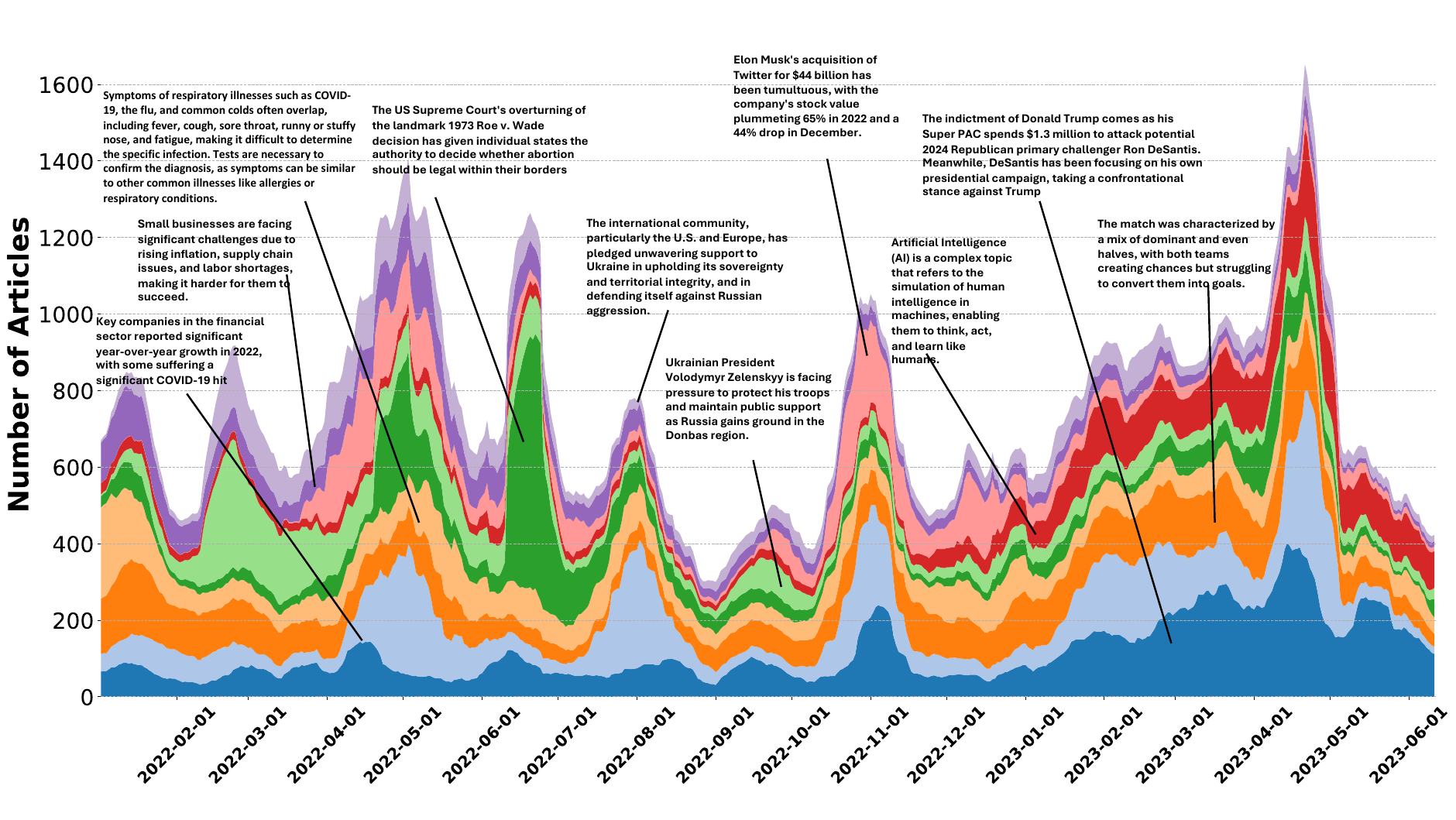}
        \vspace{-15pt}
        \caption{The most commonly discussed stories on reliable news websites labeled with their LLM-generated summaries.}
        \label{fig:mainstream-topics}
        \vspace{-5pt}
\end{figure*}

\section{Characterizing News Ecosystems\label{sec:characterize}}

Having detailed our methodology, we now characterize the ecosystem of reliable, mixed reliability, and unreliable news websites. 
Visualized in Figure~\ref{fig:mainstream-topics}, the most heavily discussed stories among our set of reliable news websites included the U.S. Republican primary (62,911~articles), business news quarterly revenue (57,453~articles), the U.S. Supreme Court's decision to overturn federal abortion rights (Roe v.\ Wade) (38,358~articles), and the Russian invasion of Ukraine (35,135~articles). Looking at the top stories spread by unreliable news websites, we observe many of the same topics, most notably one concerning the U.S. Republican primary (13,393~articles). Indeed, across all shared story clusters (91,390~stories, 62.5\%), we observe an average Pearson correlation of 0.501 between the volume of articles from our unreliable and reliable news websites. 

Beyond these shared stories, we observe a focus on corruption and government failures (9,094~articles), the U.S. Federal Bureau of Investigation's (FBI) search of President Donald Trump's Mar-a-Lago estate (7,678~articles), and the investigation into Hunter Biden's (U.S. President Joe Biden's son) laptop (7,509~articles)~\cite{Ng2022} on unreliable websites. Finally, for our set of mixed-reliability news websites, we observe a heavy focus on sports and pop culture; two of the top five topics focus on the celebrity Kardashian family and one on the footballer Cristiano Ronaldo. Mixed-reliability news volume is{also} highly correlated with the volume of stories on reliable (127,106/86.9\% shared stories with a $\rho = 0.689$ Pearson correlation for the story volumes) and unreliable new websites (91,205/62.4\% shared stories with a $\rho = 0.646$). We detail each ecosystem's stories in Appendix~\ref{sec:articles-over-time}.


Using the stance of each website toward the top 5,000~entities in our dataset, as output by our augmented {TATA} model, we further characterize the attitudes of our reliable, mixed-reliability, and unreliable news websites. To do this, we utilize PMI to determine the non-neutral stances most associated with each ecosystem (we limit this analysis to stances represented in at least 500 articles within each ecosystem to avoid spurious values; see Section~\ref{sec:appendix-pmi} for details). As seen in Table~\ref{tab:associated-keywords}, reliable news websites are more pro-CDC (Centers for Disease Control), pro-Quantum, and pro-Senate (than mixed-reliability and unreliable websites). The most distinctive stances of mixed-reliability websites concern pop culture and football (Gunnar is a Norwegian football manager and Al Nassr Football Club is a Saudi-Arabian football team). In contrast, the most distinctive stances among the unreliable news websites primarily concern the COVID-19 pandemic, with these websites distinctly opposing vaccines, Pfizer (one of the leading companies that developed a COVID-19 vaccine), and Wuhan, China (the origin of COVID-19)~\cite{Mueller2024}.

The stances between different news ecosystems are fairly distinctive. Indeed, by fitting a random forest classifier to 80\% (3,260 websites) of the websites' stance data based on their percentage for and against different entities (using 10\% of the websites as validation (408~websites) and 10\% as test data), we achieve an accuracy of 85.9\% and an AUC of 0.889 in differentiating unreliable news websites from reliable and mixed-reliability websites. This illustrates the ease of differentiating between types of websites by their stances and the ability to predict a potentially unlabeled website's reliability based on its stance towards popular news stories.

\begin{figure*}[!htbp]
 \begin{subfigure}[b]{0.49\textwidth}
        \centering
        \includegraphics[width=\textwidth]{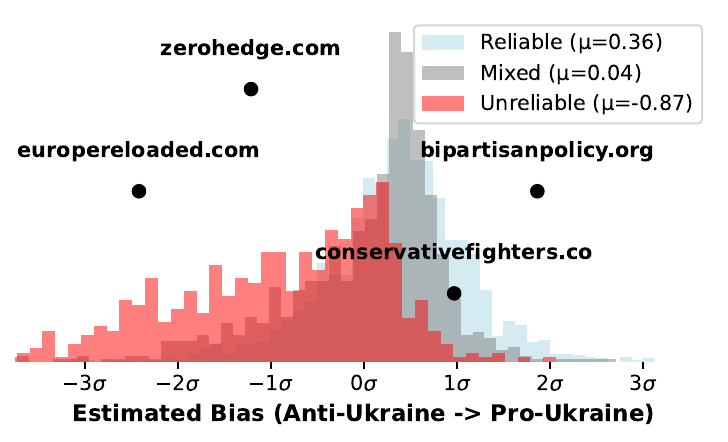}
    \end{subfigure}
    \hfill
    \begin{subfigure}[b]{0.49\textwidth}
        \centering
        \includegraphics[width=\textwidth]{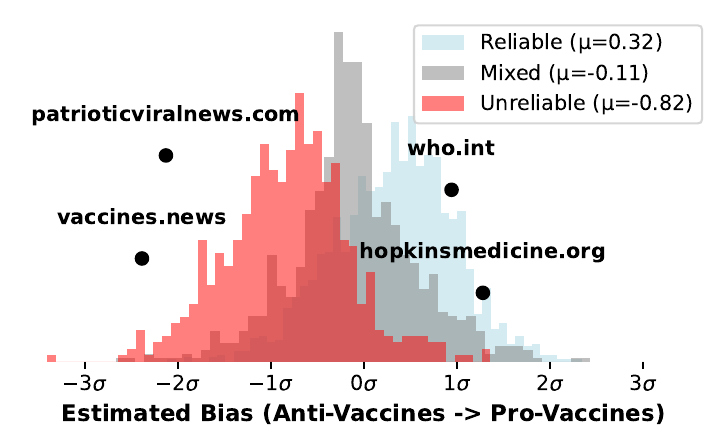}
    \end{subfigure}
    \vspace{-5pt}
    \caption{Distribution of Ukraine and Vaccine bias across unreliable, mixed-reliability, and reliable news websites estimated by  Bayesian regression models.}
    \label{fig:latents}
\end{figure*}

\vspace{2pt}
\noindent
\textbf{Bias Case Study: Ukraine and Vaccines.}\quad Beyond the most distinctive stances that each website has, to further understand the underlying attitudes within each ecosystem, we perform a case study on each website ecosystem's attitudes towards \textit{Ukraine} and \textit{Vaccines}---two of the most commonly covered topics in our dataset---using the methodology outlined in Section~\ref{sec:bias-mapping}. While this analysis specifically addresses {Ukraine} and {vaccines}, similar to how we analyzed U.S.-based political partisanship in Section~\ref{sec:bias-mapping}, this approach can be applied to any popular entity within our dataset. We additionally present analyses for {America, China, and Iran} in Appendix~\ref{sec:additional-stances}.

\begin{table}[htbp]
\centering
\small
\begin{tabular}{l|c|c}
\toprule
{Pro-Ukraine Stances} & {Coeff.} & {Std.} \\\midrule
Pro Zelenskyy & 0.378 & 0.114 \\ 
Pro Zelensky & 0.368 & 0.110 \\ 
Against Syria & 0.225 & 0.114 \\ \midrule
{Anti-Ukraine Stances}  \\\midrule
Against Zelenskiy & -0.500 & 0.111 \\
Against Biden & -0.370 & 0.103 \\
Against DHS & -0.345 & 0.119 \\
\bottomrule
\end{tabular}
\vspace{-5pt}
\caption{Stances associated with Ukraine estimated by a Bayesian regression model.\label{tab:ukraine-stances}}
\vspace{-10pt}
\end{table}

Fitting our Bayesian regression models with the stances for the remaining 4,999~entities for both {Ukraine} and {vaccines}, we map all of our news articles to a bias latent for both entities in Figure~\ref{fig:latents}. We observe that reliable news websites express higher support for {Ukraine} and {vaccines} ($\mu_{vaccine}$=0.32, $\mu_{ukraine}$=0.36), while unreliable news websites oppose both ($\mu_{vaccine}$=-0.82, $\mu_{ukraine}$=-0.87), and mixed-reliability websites in the middle ($\mu_{vaccine}=-0.11, \mu_{ukraine}=0.04$). This matches the average distribution where 31.9\% of unreliable news articles were anti-{Ukraine} and 23.8\% were anti-{vaccine}; for mixed-reliability websites, 17.8\% were anti-{Ukraine} and 8.5\% were anti-{vaccine}; and for reliable news websites 12.9\% of articles were anti-{Ukraine} and 7.1\% were anti-{vaccine}. 

Among our dataset, the news sites most anti-{Ukraine} include rt.com ($z_{ukraine}$ = -2.36), strategic-culture.org ($z_{ukraine}$ = -2.54), and southfront.org ($z_{ukraine}$ = -2.41)---three websites known for spreading Russian propaganda~\cite{RussiaPillar2020}. Some of the most pro-Ukraine websites include nationaljournal.com ($z_{ukraine}$ = +2.53), a U.S. political policy-oriented website, kyivpost.com ($z_{ukraine}$ = +0.80), a Ukrainian website, as well as a selection of NBC and ABC affiliate websites including wbaltv.com ($z_{ukraine}$ = +2.04), wvtm13.com ($z_{ukraine}$ = +2.14), and ketv.com ($z_{ukraine}$ = +2.55)~\cite{mediabias2023}. The most anti-vaccine websites are vaccineimpact.com ($z_{vaccine} = -3.41$) and pantsonfirenews.com ($z_{vaccine}$ = -2.56), both known for spreading misinformation~\cite{mediabias2023}. Conversely, the most pro-vaccine websites include Johns Hopkins ($z_{vaccine}$ = +1.28) and the World Health Organization ($z_{vaccine}$ = +0.94).

\begin{table}[t]
\centering
\small
\begin{tabular}{l|c|c}
\toprule
Pro-Vaccine Stances &  Coeff. & Std. \\
\midrule
Pro Ukraine & 0.343 & 0.111 \\ 
Pro Trans & 0.259 & 0.110 \\ 
Pro Healthcare & 0.233 & 0.093 \\\midrule
Anti-Vaccine Stances \\ \midrule
Against COVID & -0.360 & 0.144 \\
Against FDA & -0.334 & 0.122 \\
Against Pfizer-BioNTech & -0.333 & 0.143 \\
\bottomrule
\end{tabular}
\vspace{-5pt}
\caption{Stances associated with vaccines estimated by a Bayesian regression model.\label{tab:vaccine-stances} }
\vspace{-10pt}
\end{table}

Examining the stances most associated with each topic latent (Tables~\ref{tab:ukraine-stances} and~\ref{tab:vaccine-stances}), we observe that for Ukraine, this includes being pro the current president of Ukraine, Volodymyr Zelensky~\cite{Timotija2022}. Beyond this entity, we further observe the entities associated with attitudes towards toward {Ukraine} include other Ukrainian allies (\textit{e.g.}, Biden and DHS) and countries in the Global South that have battled for attention and aid following the Russian invasion of Ukraine~\cite{brosig2024war}. For the vaccine latent, we observe that website stances most associated with being pro {vaccines} have to do with being pro-health interventions like healthcare, as well as left-leaning causes like transgender rights and Ukraine~\cite{kerr2021political,kreko2024political}. In contrast, we observe that being against vaccines is associated with being against COVID (the cause of the polarization of vaccination~\cite{kerr2021political}), the US Food and Drug Administration (FDA), and Pfizer, one of the companies that developed COVID-19 vaccines~\cite{kerr2021political,PewVaccine}.

\begin{figure*}[!htbp]
 \begin{subfigure}[b]{0.24\textwidth}
        \centering
        \includegraphics[width=\textwidth]{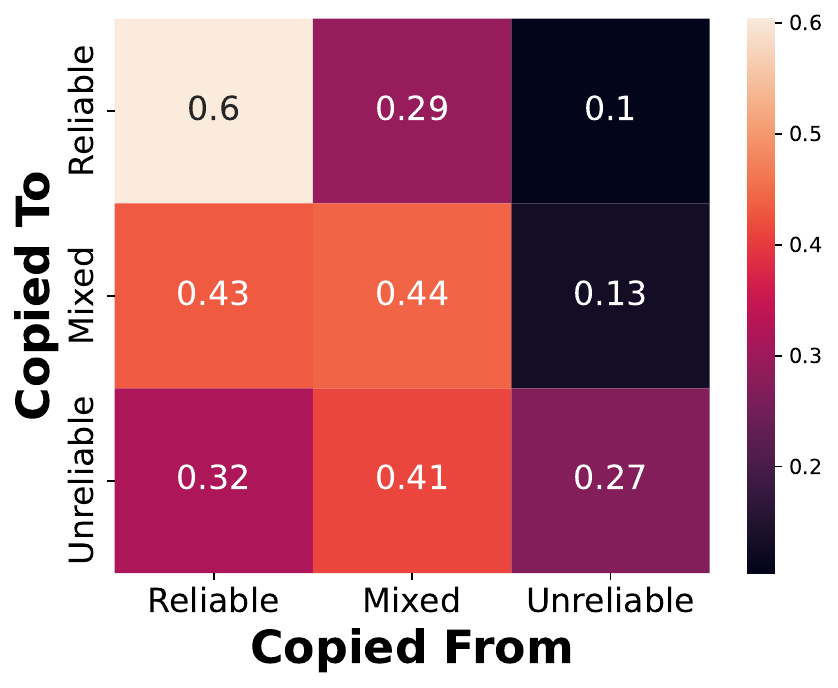}
        \caption{Copies for All Stories}
        \label{fig:copy-all}
    \end{subfigure}
    \begin{subfigure}[b]{0.24\textwidth}
        \centering
        \includegraphics[width=\textwidth]{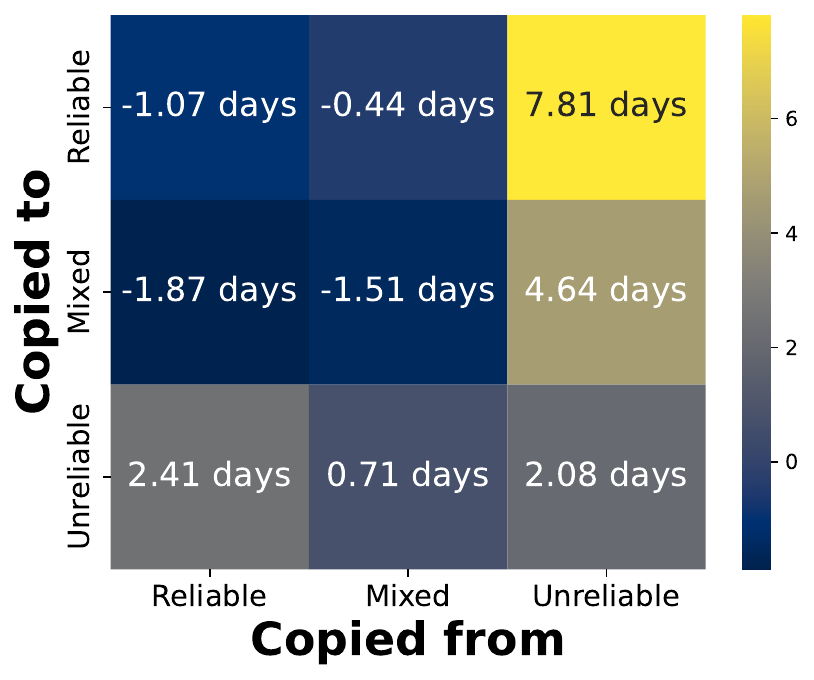}
        \caption{$\Delta$-Copy Times for All Stories}
        \label{fig:copy-time-all}
    \end{subfigure}
    \begin{subfigure}[b]{0.24\textwidth}
        \centering
        \includegraphics[width=\textwidth]{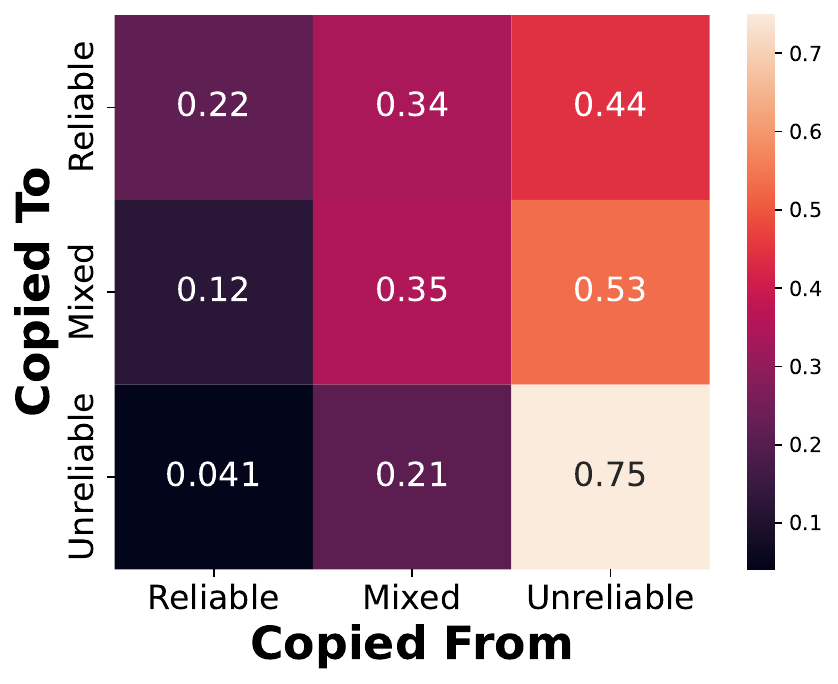}
        \caption{Copies for Unrel. Stories}
        \label{fig:copy-misinformation}
    \end{subfigure}
 \centering
    \begin{subfigure}[b]{0.24\textwidth}
        \centering
        \includegraphics[width=\textwidth]{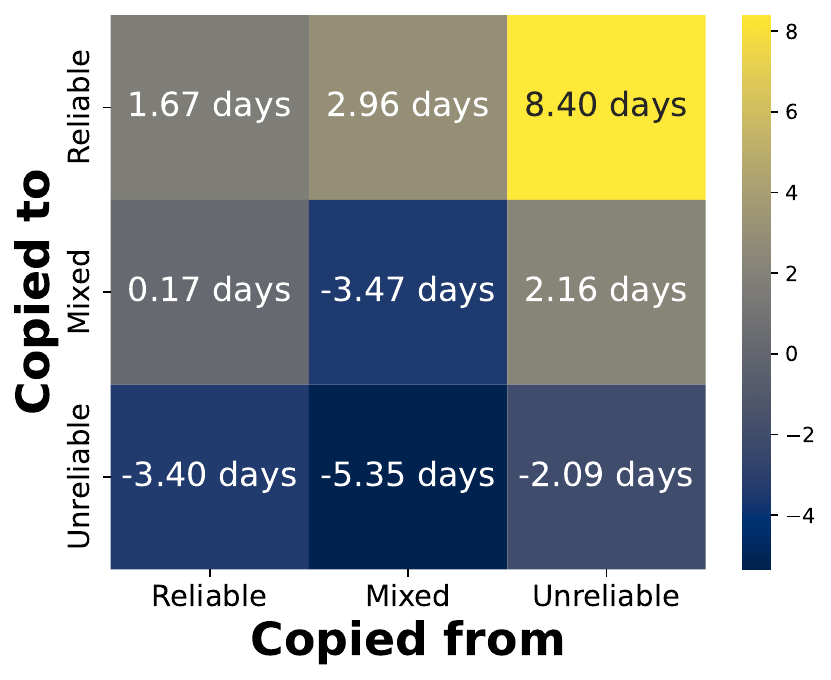}
        \caption{$\Delta$-Copy Times for Unrel. Stories}
        \label{fig:copy-time-misinformation}
    \end{subfigure}
    \caption{The percentage of each ecoystems copied stories that came from each different ecosystem as well as the change in the average time delay between website copy/reposting on the same narrative depending on the combination of news ecosystems.  }
    \label{fig:dimensions}
    \vspace{-5pt}
\end{figure*}

\section{Underlying Website Relationships}
As observed in Section~\ref{sec:characterize}, unreliable, mixed-reliability, and reliable news websites often cover the same stories simultaneously, suggesting an interdependence~\cite{starbird2018ecosystem}. To further understand these relationships, we utilize an open source version of the {NETINF}~\cite{gomez2012inferring} algorithm to infer the underlying structure and relationships amongst our sets of news websites~\cite{leskovec2009meme}. Specifically, we first run {NETINF} using all of the extracted stories within our dataset as time cascades. To determine the appropriate number of iterations to run NETINF algorithm, as in Gomez et~al.~\cite{gomez2012inferring}, we utilize the point at which the marginal gain of adding new edges plateaus (90\% of the total marginal gain). We find that margin gain reaches a plateau at 37,670~iterations in our dataset.

\vspace{2pt}
\noindent
\textbf{Ecosystem Relationships Across All News Stories.}\quad
Using the estimated number of copies between websites and the time delay between copies as found by {NETINF}, we first examine the overall relationships between ecosystems. We find that reliable and mixed-reliability news websites have a large role in introducing stories adopted by the rest of the news ecosystem. As seen in Figure~\ref{fig:copy-all}, 60\% of the news articles on reliable news websites that were copied/influenced from elsewhere came from other reliable news websites, 43\% on mixed-reliability websites came from reliable sites, and 32\% on unreliable sites came from reliable sites. Unreliable news websites had significantly less influence, with only 10\% of the stories on reliable news websites originating from unreliable news websites (13\% for mixed-reliability, 27\% for unreliable). 

Looking at the set of reliable websites that are the most common sources of copied stories throughout the entire news ecosystem, we see several popular websites including yahoo.com (1.19\%), apnews.com (0.73\%), abcnews.go.com (0.60\%), and cnn.com (0.60\%). Despite popular, reliable news websites being common sources, website popularity had only a slight Pearson correlation with their percentage of copies. Using data from the Google Chrome User Report (CrUX) from October 2022 (which Ruth et~al.~\cite{ruth2022toppling, ruth2022world} showed to be the most reliable website popularity metric), we find that for unreliable news websites copying from reliable websites, the corresponding reliable websites' popularity had a correlation of $\rho$=0.225 with the tendency of unreliable website's to copy from them ($\rho$=0.189 for reliable websites copying from reliable websites, $\rho$=0.311 for mixed-reliability news websites copying from reliable websites).

As seen in Figure~\ref{fig:copy-time-all}, reliable news websites adopt the stories of other reliable news websites more quickly (-1.07 days) compared to the average copy delay (38.4~days). Mann-Whitney U-tests indicate that these differences are all significant. This compares to a nearly +7.81 day additional delay of reliable news websites picking up the stories from unreliable news websites and -0.44 days from mixed-reliability websites. We find a similar pattern amongst mixed-reliability websites, who adopt stories from reliable news  sites (-1.87~days) more quickly than from unreliable news sites (+4.64~days). 

\begin{figure}
  \centering
  \includegraphics[width=\columnwidth]{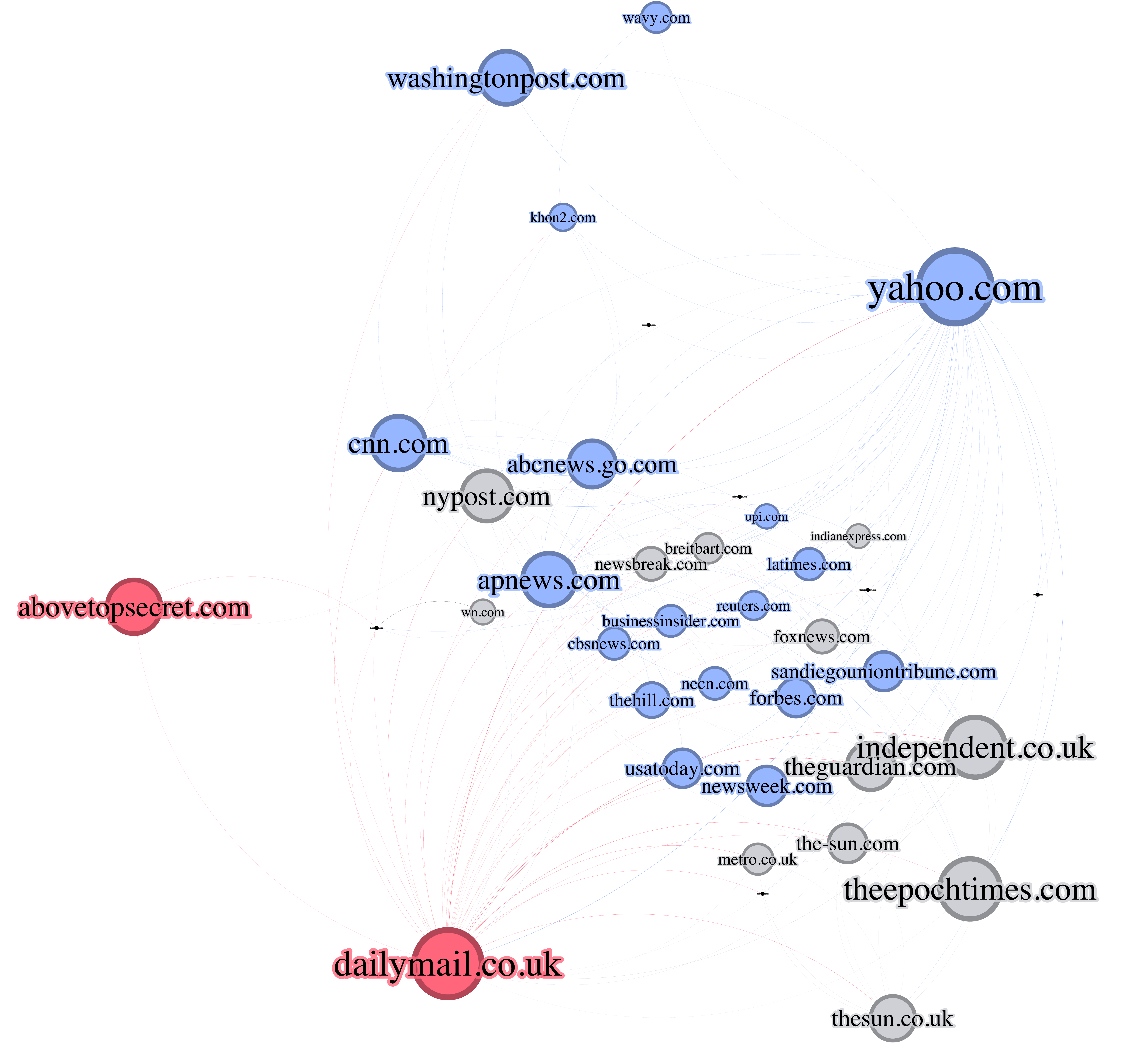}
\caption{The most influential websites and their interactions. The size of nodes is proportional to their hub centrality. Reliable news websites are colored blue, mixed-reliability websites are colored grey, and unreliable news websites are colored red.}
\label{fig:all-news-ecoystem}
\vspace{-15pt}
\end{figure}

\vspace{2pt}
\noindent
\textbf{Influence on the Full News Ecosystem. }
Having examined the website copies and rates of adoption between the different ecosystems, we next consider which websites are the most influential using the graph of the edge connections between individual news sites. Eigenvector centralities are utilized to determine the relative influence of nodes within graphs~\cite{ruhnau2000eigenvector} and, as such, we utilize this metric to understand websites' influence.  We further compute hub centralities as a metric for websites' influence in originating stories that spread to other websites (given the directionality of the arrows in our graph, this metric determines the most important websites for supplying content~\cite{kleinberg1999web}). We show the most influential sites in Table~\ref{tab:most-influential-websites} and Figure~\ref{fig:all-news-ecoystem}.

\begin{table}[h]
\centering
\small
\begin{tabular}{lll}
\toprule
{All stories} & {Hub} & {Eign.} \\ 
\midrule
yahoo.com & 0.149 & 0.111 \\
apnews.com & 0.101 & 0.092 \\
dailymail.co.uk & 0.097 & 0.099 \\
nypost.com & 0.052 & 0.076 \\
independent.co.uk & 0.049 & 0.097 \\
\bottomrule
\end{tabular}
\caption{\label{tab:most-influential-websites} Websites with the largest influence in the underlying influence graph determined by {NETINF} with all stories considered.}
\vspace{-10pt}
\end{table}

We find that website popularity is correlated with the relative influence of websites within the news ecosystems (when looking at all news sites compared to only reliable sites in the last section). Again using website popularity data from the Google Chrome User Report (CrUX), we find that a website's eigenvector centrality/influence has a Spearman correlation of $\rho =$ 0.571 (0.396 for hub centrality) with that website's popularity rank. 
Despite making up 24.6\% of the news websites in our dataset, unreliable news websites do not make up a proportional percentage amongst the most influential news websites. Directly comparing the eigenvector centralities of the reliable news websites to those of the unreliable news websites, we find that reliable news websites are significantly more influential in this ecosystem than unreliable news websites (Cohen's~D~=~0.64, p-value~$\approx$~0),\footnote{The p-value is computed using the Mann-Whitney U-test.} with mixed reliability websites having comparable influence to reliable ones (no significant difference through Mann Whitney U-test). In terms of origination (hub centralities), we observe a slightly different trend with mixed-reliability websites having slightly more influence in originating stories compared to reliable news websites (Cohen's~D~=~0.04, p-value~$\approx$~0) and unreliable news websites (Cohen's~D~=~0.05, p-value~$\approx$~0).

\vspace{2pt}
\noindent
\textbf{Stories Spread by Unreliable News Websites. }\quad
To identify the websites most inflential in spreading potentially unreliable stories, we run the {NETINF} algorithm on the set of 6,762~news stories where unreliable news websites posted the plurality of articles about those stories. Again applying the same methodology for identifying the appropriate number of edges to add, the marginal gain plateaus at 16,196 edges. The most popular story amongst these clusters was about government censorship and control (9,094~articles) summarized as: \textit{
There is censorship, propaganda, and
government control in the US. Cancel
culture is a form of censorship, and that
government-funded media outlets can
exercise control over editorial content.
The text also warns about the influence
of the ``Deep State'' and far-left
communists in US institutions,
including the government, media,
education, and Big Business.}\looseness=-1

\begin{table}[t]
\centering
\small
\begin{tabular}{lr}
\toprule
{Reliable} & {Propor.}  \\ \midrule
dailymail.co.uk & 0.140 \\
abovetopsecret.com & 0.039 \\
ussanews.com & 0.035 \\
\midrule
{Mixed} & \\ \midrule
dailymail.co.uk & 0.062 \\
thegatewaypundit.com & 0.030 \\
ussanews.com & 0.027 \\
\midrule
{Unreliable} & \\ \midrule
naturalnews.com & 0.025 \\
ussanews.com & 0.021 \\
theburningplatform.com & 0.020 \\
\bottomrule
\end{tabular}
\vspace{-5pt}
\caption{\label{tab:common-misinfo-sources} Websites that are the most common source of unreliable news stories for each news ecosystem.}
\end{table}

As expected, given how we narrow our set of stories, as seen in Figure~\ref{fig:copy-misinformation}, relative to all news stories, unreliable news websites had significantly more influence in originating potentially unreliable content. For example, while for all stories, reliable news websites sourced less than 10\% of all of their stories from unreliable news websites, within this specific set of news stories, the figure was 44\%. Similarly, for mixed-reliability websites, this percentage increased from 13\% to 53\%. Furthermore, we find that unreliable websites source the majority of their influenced or copied stories from other unreliable news websites, at a rate of 75\%. Looking at the set of websites that are the common source for other sites to copy from (Table~\ref{tab:common-misinfo-sources}), we find a heavy reliance on dailymail.co.uk, a United Kingdom-based tabloid that Media-Bias/Fact-Check describes as having ``low'' factual reporting due to ``numerous failed fact checks and poor information sourcing.'' We also find that ussanews.com, described by Media-Bias/Fact-Check as promoting ``entirely false, so-called facts,'' was a common source of unreliable news stories.

For this selection of news stories predominately published by unreliable news websites, comparing the copy times of these stories in Figure~\ref{fig:copy-time-misinformation} to those in Figure~\ref{fig:copy-time-all}, we find that reliable news websites are slower to adopt the stories, regardless of from which news ecosystem the story originated. We thus observe a reticence amongst our reliable news websites to report on the news stories primarily spread by unreliable news outlets.  However, we find that for mixed-reliability websites, if the news story began amongst other mixed-reliability news outlets, these news outlets are faster to adopt the story (-3.47~days). We further observe that unreliable news websites are the fastest at picking up these news stories compared, picking them up quicker if they initially came from a mixed-reliability (-5.35 days) or reliable news website (-3.40 days).

\begin{table}[h!]
\centering
\small
\selectfont
\setlength\tabcolsep{4pt}

\begin{tabular}{lll}
\toprule
{Predom. Unreliable News Stories} & {Hub} & {Eign.} \\ 
\midrule
thegatewaypundit.com & 0.129 & 0.109 \\
dailymail.co.uk & 0.075 & 0.125 \\
theburningplatform.com & 0.060 & 0.103 \\
\bottomrule
\end{tabular}
\vspace{-5pt}
\caption{\label{tab:predominant-misinfo} Websites with the largest influence in the underlying influence graph for stories predominated spread by unreliable websites.}
\end{table}

\begin{figure}
  \centering
  \includegraphics[width=1\columnwidth]{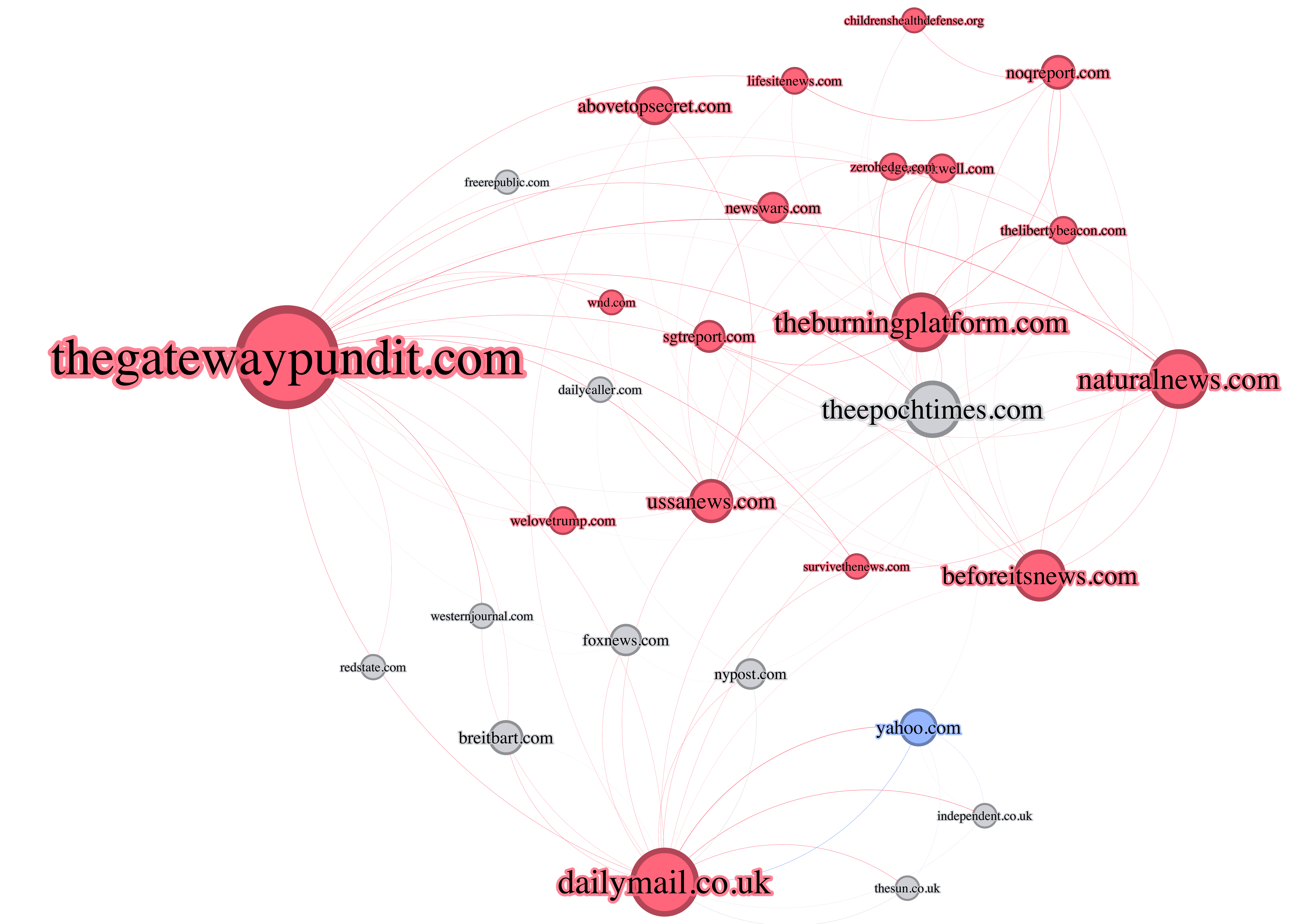}
\caption{Most influential websites and their interaction for stories that are predominantly spread by unreliable news websites. Nodes sizes are proportional to their hub centralities.}
\label{fig:misinformation-news-ecosystem}
\vspace{-5pt}
\end{figure}

\noindent
\textbf{Influence in the Unreliable News Ecosystem.}\quad
To understand which websites are the most influential in the unreliable news ecosystem, we utilize the eigenvector centrality of each website in the resultant graph created by running {NETINF} on our set of predominantly unreliable news stories (Table~\ref{tab:predominant-misinfo}). For this ecosystem, we find the popularity of websites is only slightly correlated with eigenvector centrality/influence ($\rho$= 0.158) and hub centrality ($\rho$= 0.175). Examining the set of websites that are most prominent within the unreliable news ecosystem (Table~\ref{tab:predominant-misinfo} and Figure~\ref{fig:misinformation-news-ecosystem}), we find that many well-documented websites known for spreading unreliable information are among the most prominent, including theepochtimes.com, dailymail.co.uk, and thegatewaypundit.com~\cite{starbird2018ecosystem}. 

Comparing the eigenvector centralities of the unreliable news websites to those of the authentic news websites, we find that unreliable news websites are more influential within this ecosystem (Cohen's~D~=~0.219, p-value~$< 0.001$), but that unreliable and mixed-reliability websites had comparable influence (no significant difference via the Mann-Whitney U-test). However, most notably, we observe that among the top influencers within this ecosystem are the reliable news website, Yahoo News, and the mixed-reliability Fox News (not shown in the table). Yahoo News primarily serves as a news aggregator, gathering reports from various sources including Fox News, the BBC, and Reuters~\cite{yahoo2023}. Given its role as an aggregator, Yahoo News appears to have a prominent role in disseminating current events that are reported by other outlets. Classified as a mixed-reliability news source, Fox News has been widely commented upon for its role in disseminating hyperpartisan news and misinformation~\cite{hoewe2020role,Bauder2023,bauer2022fox}.

\vspace{2pt}
\noindent
\textbf{Case Study: News Website Coordination.}\quad
To identify potential coordination among our websites, we utilize {NETINF} to discern the relationships between websites involved in stories predominantly published articles spread by unreliable and mixed-reliability news websites (encompassing 40,325~news stories). After running the {NETINF} algorithm, we further clustered the resulting graph using the Louvain clustering algorithm~\cite{de2011generalized}. Qualitatively, the largest of these clusters comprised 885~relatively mainstream and tabloid websites that report on general news (\textit{e.g.}, wpxi.com, nbc29.com, wvva.com), with the top stories concerning the Kardashians (\textit{Keywords: Kourtney, Kardashian, Travis, Khloe, Barker}). The second largest cluster consisted of 492~locally-oriented news websites (\textit{e.g.}, cbs4local.com, idahostatejournal.com), where the top stories focused on immigration (\textit{Migrant, Border, Patrol, Customs, Smuggling}) and the US Constitution (\textit{Constitution, Oath, Amendment, Constitutional}). Finally, the third largest cluster (\textit{Crore, Yoy, Profit, FY23, Quarter}) included 334 international websites (\textit{e.g.}, sputniknews.com, alarabiya.net), where the top story involved international companies' profits.

Most notably among our clusters was a set of 338~websites, all with seemingly innocuous names such as southindynews.com and northalaskanews.com, which appeared to be dedicated to local news. Upon further investigation through querying WHOIS, we discovered that each of these websites was registered by the domain registrar Epik, Inc., a popular provider for misinformation and online hate sites~\cite{haninfrastructure}. We find that this set of 338~ostensibly local websites is owned and operated by the same entity, Metric Media LLC, which produces algorithmically generated content and promotes right-wing views~\cite{metric2023}. Indeed, using our mapping of websites to their respective political partisanship, we found that despite these websites rarely writing articles about Republicans or Democrats, they have an average partisanship $\mu_{politics}=0.22$, indicating a slight right-leaning bias, with 88.2\% of these websites classified as right-leaning. These websites largely repeat the same text including articles promoting herd immunity from COVID-19 in the United States: \textit{More than 50 percent of US citizens are considered fully vaccinated against COVID-19, nearing the target for ``herd immunity'' Herd immunity happens when enough of the population has become immune to the virus from the previous infection that it effectively protects those who are not immune.}




\section{Propaganda and Slanted Influence Networks}
As seen in the last section, news sites, regardless of their factual reliability, often report on the same stories, with unreliable news websites in select cases influencing both reliable and mixed-reliability news platforms. Furthermore, while reliable and mixed-reliability news websites predominantly adopt stories from other reliable and mixed-reliability sources (Figure~\ref{fig:copy-all}), for topics primarily spread by unreliable news websites, these specious sources often act as the originators of the content (Figure~\ref{fig:copy-misinformation}). Within this vein, tracking the spread of unreliable news and propaganda and determining which sources are most effective at seeding these stories into the mainstream media is critical for fact-checkers, journalists, and researchers~\cite{starbird2018ecosystem,hanley2023specious}. To this effect, in this section, we utilize our system to map out and understand the sites originating and spreading specific propaganda and influence campaigns.

\begin{figure}
        \centering
        \includegraphics[width=0.35\textwidth]{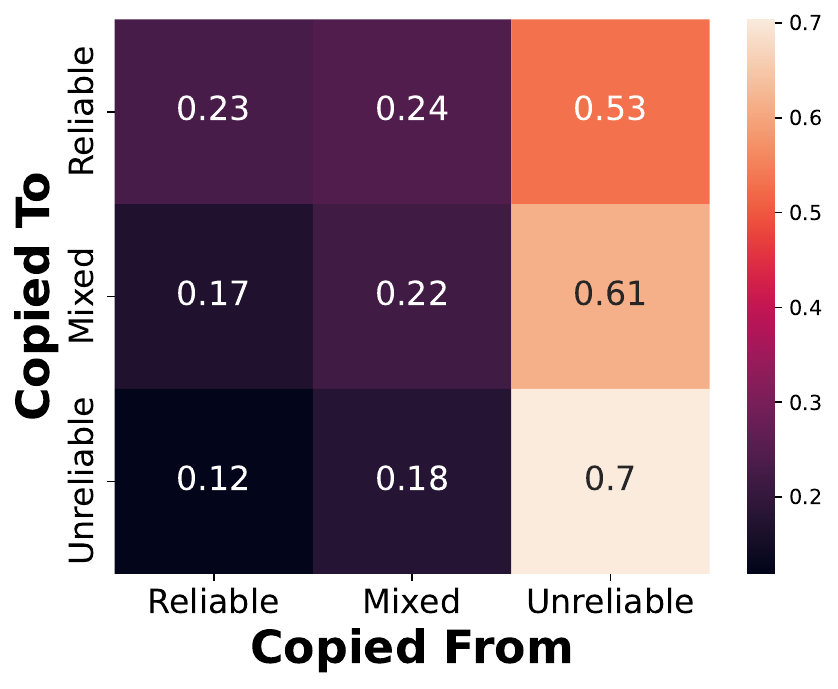}
    \vspace{-5pt}
       \caption{Anti-Ukraine Copy Matrix. }
    \vspace{-5pt}
        \label{fig:anti-ukriane-copy}

\end{figure}

To map the influence networks targeting specific entities (either positively or negatively), we gather news articles and the associated sites that exhibit a particular valence towards a given entity (\textit{e.g.}, anti-vaccine articles). Upon gathering this subset of news articles, we run the {NETINF} algorithm over these cascades of news article clusters with specific stances. We subsequently perform network analysis using eigenvector centrality and hub centrality, as discussed in the previous section, to identify the most prominent and influential sites promoting a given stance towards a particular subject. By further examining day-to-day increases in news stories with specific stances and comparing their spread in reliable and unreliable news ecosystems, we further document \emph{new} individual stories meant to spread particular views or stances.

This programmatic approach can help identify stories that are receiving renewed focus from unreliable news websites and which websites are influential in propagating stances towards entities of interest in a particularly damaging manner, thereby facilitating the identification and mitigation of misinformation~\cite{rajdev2015fake, wu2019misinformation, saeed2022trollmagnifier,hanley2023specious}. To illustrate, we perform this analysis for anti-vaccine and anti-Ukraine news stories.

\begin{table}[ht]
\centering
\small
\begin{tabular}{lll}
\toprule
{Anti-Ukraine} & {Hub} & {Eign.} \\
\midrule
rt.com                & 0.155 & 0.210 \\
sputniknews.com       & 0.073 & 0.129 \\
news-front.info       & 0.054 & 0.163 \\
\midrule
{Anti-Vaccine} & & \\ \midrule
naturalnews.com              & 0.141 & 0.179 \\
theepochtimes.com            & 0.086 & 0.196 \\
vaccines.news                & 0.049 & 0.170 \\
\bottomrule
\end{tabular}
\vspace{-5pt}
\caption{\label{tab:most-influential-websites-anti} Websites with the largest influence in the underlying influence graph of anti-vaccine and anti-Ukraine news.}

\end{table}

\begin{figure}
 \centering
        \centering
        \includegraphics[width=\columnwidth]{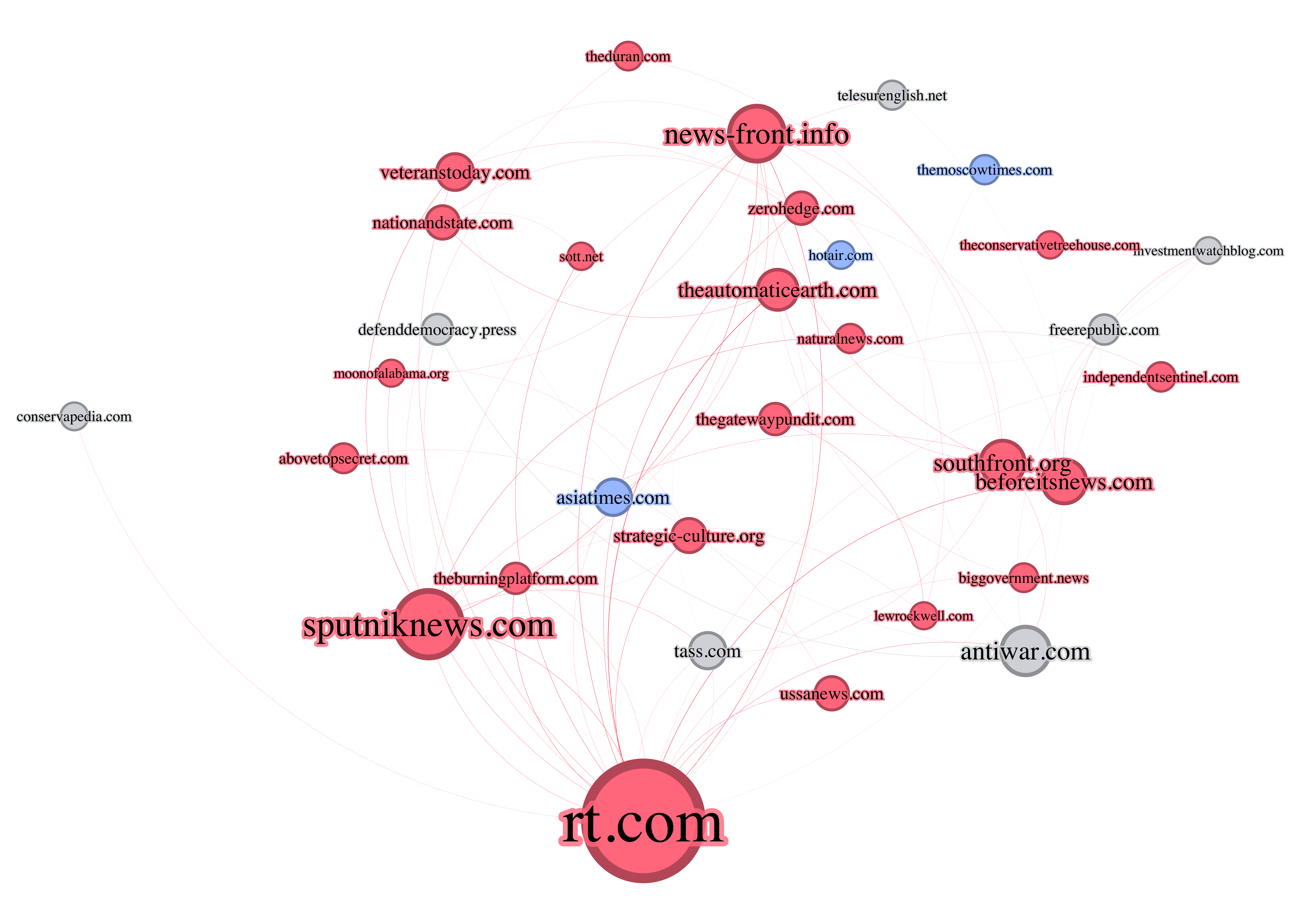}
        \vspace{-20pt}
        \caption{Anti-Ukraine Influence Network determined by the NETINF algorithm. The nodes' sizes are proportional to their hub centralities.}
        \label{fig:anti-ukraine-graph}
        \vspace{-5pt}
\end{figure}

\vspace{2pt}
\noindent
\textbf{Anti-Ukraine Messaging.}\quad As seen in Figure~\ref{fig:anti-ukraine-graph} and Table~\ref{tab:most-influential-websites-anti}, the most prominent anti-Ukrainian news websites during our study included well-known Russian propaganda websites such as Russia Today (RT), Sputnik News, and NewsFront~\cite{RussiaPillar2020}. Beyond known Russian propaganda sites, we also find that antiwar.com, described as a ``libertarian non-interventionist website''~\cite{mediabias2023}, is one of the most prominent websites in spreading anti-Ukrainian content. Altogether, as seen in Figure~\ref{fig:anti-ukriane-copy}, unreliable news websites largely supply the majority of the stories used across the entire news ecosystem, with reliable websites copying 53\% of anti-Ukrainian stories from unreliable outlets, mixed-reliability websites copying 61\%, and unreliable sites 70\%.

The most common story pushed in this ecosystem of websites concerned justifications for Russia's invasion of Ukraine, with one Russia Today article writing~\cite{RT2022}: \textit{Moscow attacked the neighboring state in late February, following Ukraine’s failure to implement the terms of the Minsk agreements signed in 2014, and Russia’s eventual recognition of the Donbass republics of Donetsk and Lugansk}. Further, the anti-Ukrainian news story that received the largest increase in relative popularity among our unreliable news websites in the last week of our study (June 25 to July 1, 2023) featured a series of articles with the keywords \textit{Ukraine, MacGregor, Douglas, Colonel, Zelensky}, showing a ratio of 9 articles in the unreliable news ecosystem for every 1 article in the reliable news ecosystem. This story, which predominantly spread within the unreliable news ecosystem, concerned an interview with retired US Colonel Douglas MacGregor suggesting that the war between Ukraine and Russia was unwinnable and that Ukrainian President Zelensky was a puppet of Western powers: \textit{\textit{``The war is really over for the Ukrainians. I don’t see anything heroic about the man. And I think the most heroic thing he can do right now is to come to terms with reality," retired Army Colonel Douglas MacGregor told Fox Business News. "I think Zelensky is a puppet, and he is putting huge numbers of his own population in unnecessary risk," he said.}} The website that spread this story the most was paulcraigroberts.org ($z_{ukraine}$=-3.59, 3 articles).


Beyond the set of unreliable news websites spreading anti-Ukrainian messaging, we further observe several international news websites including asiatimes.com ($z_{ukraine}$ =-1.13), themoscowtimes.com ($z_{ukraine}$ =-0.67), and the right-leaning website hotair.com ($z_{ukraine}$ =-1.21) as purveyors of influential anti-Ukrainian content in this ecosystem. Based on the inward-weighted edges, the most influenced mainstream news website in this ecosystem was haaretz.com, an Israeli outlet ($z_{ukraine}$ =+0.002 for Ukraine bias), and the most influenced mixed-reliability news website was salon.com ($z_{ukraine}$ = -0.056), a US-based left-leaning news outlet. We thus observe that even relatively neutral and pro-Ukrainian websites can be potentially influenced by anti-Ukrainian news articles.

\begin{figure}
        \centering
\includegraphics[width=0.35\textwidth]{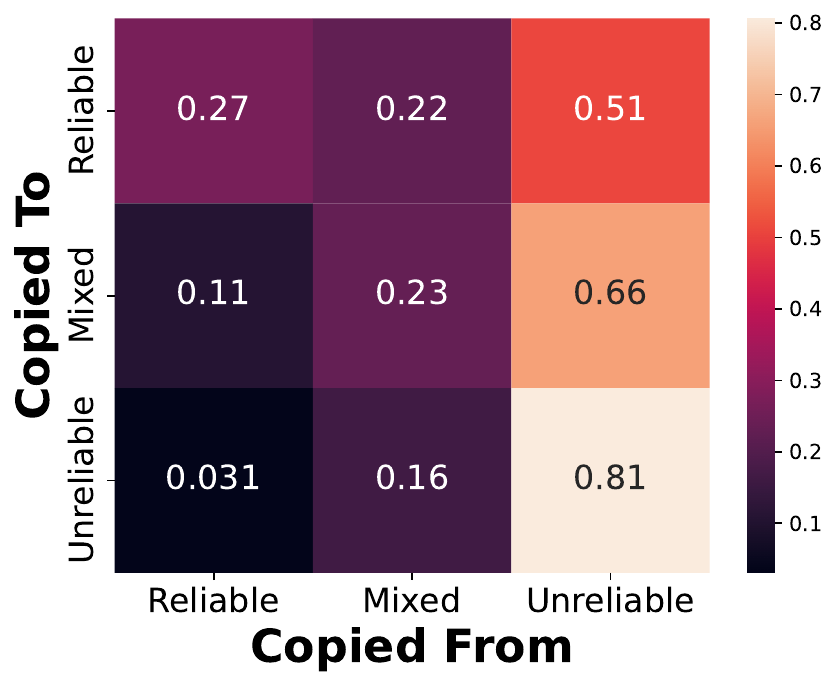}
    \vspace{-5pt}
       \caption{Anti-Vaccine Copy Matrix}
    \label{fig:anti-vaccine-all}
    \vspace{-5pt}
\end{figure}

\vspace{2pt}
\noindent
\textbf{Anti-Vaccine Messaging.}\quad As seen in Figure~\ref{fig:anti-vaccine-all} and Table~\ref{tab:most-influential-websites-anti}, the largest source of anti-vaccine stories was naturalnews.com, while the most influential anti-vaccine website was theepochtimes.com, both known for spreading anti-vaccine misinformation~\cite{mediabias2023,FactCheck2023}. As with anti-Ukraine stories, we observe that each website category predominantly sourced their content from unreliable news websites: 51\% for reliable news websites, 66\% for mixed-reliability news websites, and 81\% for unreliable news websites. In addition to the theepochtimes.com and naturalnews.com, we find that childrenshealthdefense.org, a website associated with former presidential candidate Robert F. Kennedy Jr., had a major influence on spreading anti-vaccine content, including one article suggesting that a vaccine was not as safe as the US Food and Drug Administration claimed~\cite{RFK2022}. 

\begin{figure}
 \centering
\centering
\includegraphics[width=\columnwidth]{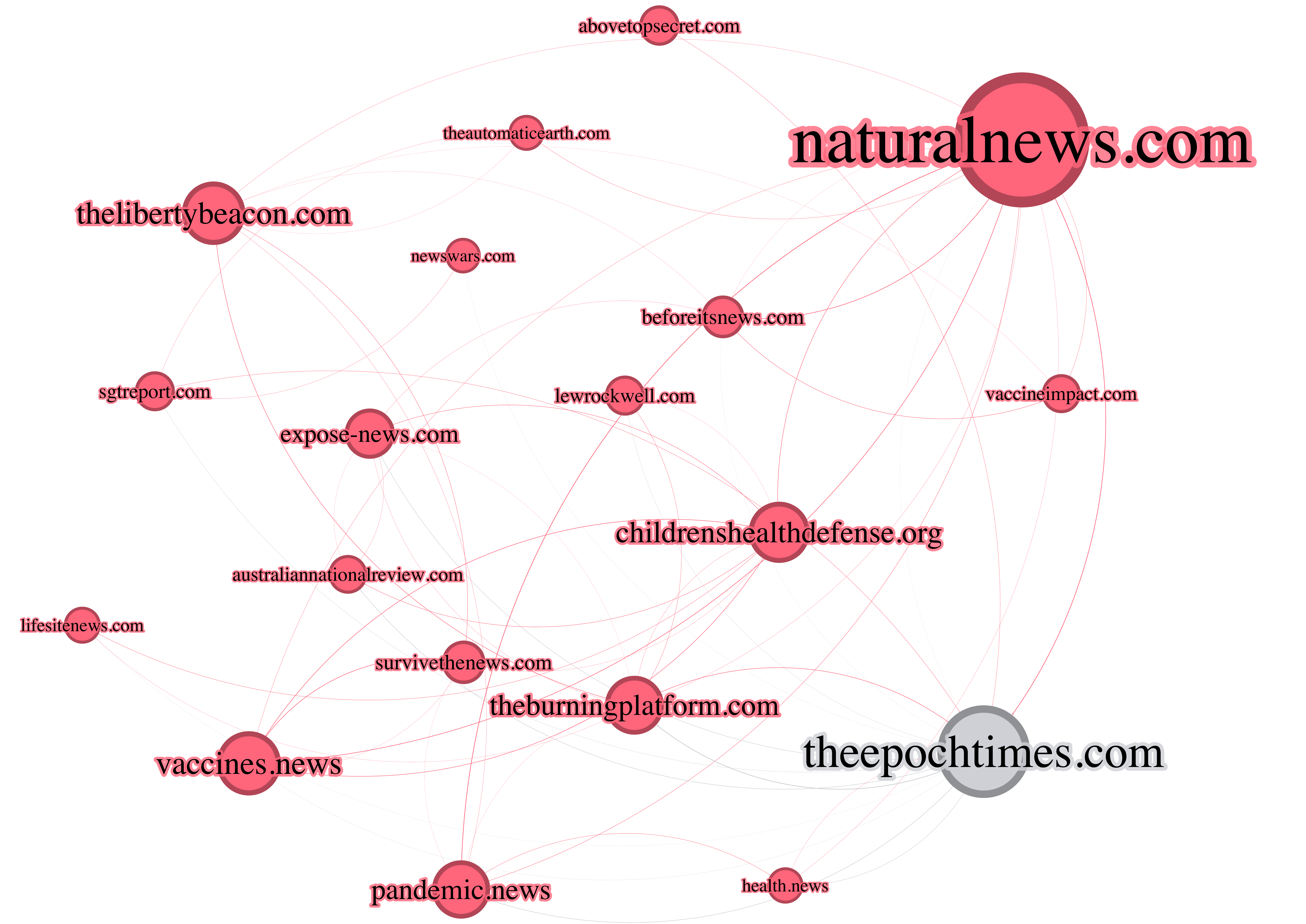}
\vspace{3pt}
        \caption{Anti-Vaccine Influence Network determined by NETINF\@. The nodes' sizes are proportional to their hub centralities.}
        \label{fig:anti-vaccine-graph}
        \vspace{-5pt}
\end{figure}

The most prominent-anti-vaccine story in terms of article volume raised concerns about children receiving COVID-19 vaccines, as highlighted by childrenshealthdefense.org~\cite{Mercola2022}: \textit{Pfizer, at the urging of federal health officials, is hustling to get infants and toddlers injected with experimental COVID vaccines}. The story that saw the largest relative increase in news articles (14 articles in the unreliable news ecosystem for every 1 in the reliable news ecosystem) was one with the keywords \textit{Pfizer, Batch, Danish, Bnt162b2, Adverse}. This story concerned Danish scientists ostensibly discovering that batches of Pfizer vaccines were actually placebos: \textit{\textit{The Danish scientists uncovered “compelling evidence” that a significant percentage of the batches distributed in the EU likely consisted of “placebos and non-placebos,” prompting the researchers to call for further investigation.}}. The top sites that spread this narrative were sgtreport.com ($z_{vaccine}$=-1.31 for vaccine-bias), theautomaticearth.com ($z_{vaccine}$ =-1.08), and theburningplatform.com ($z_{vaccine}$=-1.86) with two articles each. 

We find that the reliable news website most influenced (by the weighted in-degree within the resulting {NETINF} graph) was sciencebasedmedicine.org ($z_{vaccine}$= +0.06), which frequently reports on and quotes anti-vaccine information~\cite{Vaccine2024}, detected by our system. Additionally, the most influenced mixed-reliability website (besides theepochtimes.com) was thelibertyloft.com ($z_{vaccine}$=-0.81), a right-leaning website that Media-Bias/Fact-Check has identified as spreading COVID-19 related misinformation~\cite{mediabias2023}.


\section{Limitations and Future Work}

Our work shows the promise of mapping the global trajectories of news stories and the takes of news sites towards specific entities. However, we the emphasize the complexity of the global news ecosystem and the considerable future work that remains to understand how information travels online. Below, we discuss the limitations of our work and potential future research directions. 

\vspace{2pt}
\noindent
\textbf{English-Language Websites.}\quad
Our work is limited to English-language news articles and focuses predominantly on US, UK, and Australian websites. As a result, our analysis of the spread of particular stories is limited largely to the English speaking world and could miss other sources of news (\textit{i.e.}, a Russian-language website for example may be more influential in spreading pro-Russian propaganda than the websites in our dataset). This restriction is largely due to our use of PMI for identifying keywords for stance detection amongst our story clusters, which does not directly work in a multilingual setting. Similarly, we currently lack highly accurate multilingual topic-agnostic stance-detection models~\cite{hardalov2022few,hanley2023tata}. We encourage future work to consider how to semantically map both news topics and stances towards them in multilingual settings, as well as to consider how to source news content from websites in additional languages.

\vspace{2pt}
\noindent
\textbf{Automated Fact-Checking of Narratives.}\quad As previously noted, we do not fact check individual news stories, which we argue is a journalistic task beyond the scope of our automated approach. While our system can be utilized to uncover networks of websites pushing potentially unreliable news narratives allowing journalists to prioritize which stories need to be fact-checked by their relative spread, these stories still require human investigation to determine their veracity. However, we note that for stories that have already fact-checked on  reputable websites, it may be possible to incorporate the approaches of Hanley et al.~\cite{hanley2023specious}, Zhou et~al.~\cite{zhou2024correcting}, and others to automatically label particular stories. Hanley et~al.'s approach involves gathering fact-checks from reputable sources and using a DeBERTa-based model~\cite{he2022debertav3}  to identify unreliable news stories that directly contradict these fact-checks~\cite{hanley2023specious}. In a similar fashion, Zhou et~al.'s~\cite{zhou2024correcting} approach involves using a LLM agent and Google Search to identify which unreliable news stories contradict fact-checks. 

\vspace{2pt}
\noindent
\textbf{Ephemeral Unreliable News Websites.}\quad
Factually unreliable news tend to be ephemeral~\cite{hounsel2020identifying,hanley2022no,dahlke2023quantifying,moore2023exposure}, often only being active long enough to spread misinformation to other platforms before shutting down themselves. As such, finding news sites as soon as they come online is critical long term. We note that while our current system relies on previously curated lists of websites, it can easily incorporate new websites as they appear (\textit{e.g.}, using the methods outlined by Hounsel et~al.~\cite{hounsel2020identifying} for identifying new unreliable news websites based on their domain registration and network infrastructure characteristics). This inclusion would enable our system to surface potentially unreliable news stories that have not spread onto more popular websites. It would also potentially enable uncovering malicious Doppelgänger sites that masquerade as ordinary local websites but that actually spread propaganda as soon as they come online~\cite{rf-impersonate,doj-doppelganger,eu-doppel} similar to past work that has detected phishing and malware domains~\cite{antonakakis2010building,antonakakis2011detecting}.
\section{Discussion and Conclusion}
In this work, we investigated the spread and stance of news stories across 4,076~news websites from January 1, 2022, to July 1, 2023. Our approach, which advances previous methodologies for understanding news flows by incorporating stance into how we track semantic narratives, allows us to track stories across a mix of reliable, mixed-reliability, and unreliable news websites. (Neglecting stance in understanding the spread of narratives, while helpful for examining a singular ecosystem ~\cite{starbird2018ecosystem,hanley2023specious}, would likely led to misrepresentations of the interactions between different types of websites.)

Our work demonstrates the key role that reliable news platforms play in dictating the stories covered by the entire news ecosystem. These popular and largely factual websites maintain the largest degree of influence on the broader news ecosystem (Figure~\ref{tab:most-influential-websites}) and are the source of much content on mixed-reliability and unreliable websites (Figure~\ref{fig:dimensions}). To understand which stories unreliable websites will spin or contort, researchers should consider reliable outlets as agenda-setters~\cite{erbring1980front,boynton2016agenda,mccombs2005agenda}. However, we simultaneously highlight that while a minimum of 62.4\% of stories are shared between different types of news websites (Section~\ref{sec:characterize}), different ecosystems often have distinctive attitudes towards stories. For example, using our analysis, inline with prior work~\cite{garrett2021conservatives,ecker2019political}, we show that current lists of unreliable websites, among other biases, tend to be more conservative and have distinctive biases against COVID-19 vaccines and Pfizer (Table~\ref{tab:associated-keywords}), informing which larger topics may particularly need additional fact-checking. We also find that biased coverage of particular entities (\textit{e.g.}, Ukraine, or vaccines) that otherwise reliable news websites produce are often sourced from unreliable news sites.

Finally, our work demonstrates how, by analyzing the stance of articles towards specific topics, we can uncover and understand influence networks directed at specific entities, facilitating the tracking of propaganda (\textit{e.g.}, anti-Ukraine) or misinformation (\textit{e.g.}, anti-vaccine) within the news ecosystem. This method also aids in identifying which otherwise reliable news sources may be influenced by disinformation and propaganda campaigns. Our approach, which considers the context of authentic and mainstream websites, provides a valuable tool for identifying dubious networks of websites spreading particular types of slanted information, which we argue can assist fact-checkers, journalists, and researchers in better understanding potential online misinformation. 

We hope that our work encourages further quantitative analysis of the distributed news ecosystem, particular as social media platforms become more opaque to researchers. Prior security research has uncovered weaknesses and attacks through large-scale analysis (\emph{e.g.},~\cite{meiklejohn2013fistful, heninger2012mining, durumeric2013analysis, thomas2015ad, sundara2020censored, durumeric2015neither, sundara2023global, acar2014web}), and we argue that there is significant potential for future work within the security community on understanding attacks against and strengthening the resilience of the news ecosystem.



 \section*{Acknowledgments} 
This work was supported in part by the NSF Graduate Fellowship DGE-1656518, a Meta Ph.D. Fellowship, and a Sloan Research Fellowship. Any opinions, findings, and conclusions or recommendations expressed in this paper are those of the authors and do not necessarily reflect the views of the National Science Foundation or other funding agencies.

\clearpage

\section*{Ethical Considerations}

Trustworthy news media is fundamental to a democratic society. Previously, false information has incited real-world violence and had major consequences on public health and elections. Disinformation and propaganda are \emph{attacks}, and it behooves the security community to understand how these attacks are conducted and how to build better defenses against them. Advances in this space help both citizens and news outlets themselves, who regularly fact-check articles. At the same time, like all active measurements, web crawling and programmatic analysis of online content have potential ethical ramifications that we must carefully consider.

Our work collects only publicly available news content in line with prior work (\textit{e.g.}, \cite{singrodia2019review,hanley2022golden,smith2013dirt}). We follow best practices when scraping websites by slowly collecting content over time to reduce load. Our scraping also includes built-in safety mechanisms to prevent making requests more often than once every 10 seconds. We never attempt to access any privileged or private data but rather focus on public stories that are linked from news platforms' public homepages.

We also adhere to the best practices set forth for conducting active Internet measurements~\cite{acar2014web, durumeric2013zmap, durumeric2024ten}. The servers we use for collecting content are identified as part of a research study through WHOIS, reverse DNS, and informational websites that indicate how to reach the researchers. Our IT and security teams are also informed about how to route any questions, requests, or complaints to our team. We received no requests to opt out of our data collection during our study.

Our study does not generate any new content or redistribute existing content. Instead, we analyze how context spreads. We emphasize that while we utilize labels of individual websites as \textit{unreliable} or \textit{mixed-reliability} from Media-Bias/Fact-Check~\cite{mediabias2023} and on existing previously-curated lists, this does not necessarily mean that every news story spread by these websites is misinformation. Many unreliable news websites report factual information~\cite{starbird2018ecosystem}, and at times, otherwise reliable websites may mistakenly report incorrect information. We only label stories that have been previously and individually expertly labeled as \textit{misinformation}.

\section*{Open Science} We are committed to sharing our data with other researchers at academic or non-profit institutions seeking to conduct future work or re-implement our approach. We will publicly release the weights and the code for the models used in this study. Additionally, we will supply the URLs of crawled news stories used in this study upon request.

\clearpage



\bibliographystyle{plain}
\bibliography{paper}
%

\newpage
\appendix

\clearpage

\section{Article Preprocessing\label{sec:appendix-pre-proces}}
After collecting each page's HMTL, we then parse the content to extract the news article text and publication date using the Python libraries \texttt{newspaper3k} and \texttt{htmldate}. We subsequently remove any leftover boilerplate language (\textit{i.e.}, navigation links, headers, and footers) from the text using the \texttt{justext} Python library and remove any non-English articles based on labels provided by the Python \texttt{langdetect} library.

To prepare our news article data for embedding, we first remove any URLs, emojis, and HTML tags from the text. Then, in line with prior work, we subsequently divide these our articles in constituent \textit{passages} with at most 100~words~\cite{piktus2021web,hanley2022happenstance,hanley2022partial}.  Specifically, after first separating articles into
different paragraphs by splitting their text on (\escape{n}) or tab (\escape{t})
characters and breaking apart each paragraph into its constituent sentences~\cite{bird2009natural}, we add sentences from a single paragraph to create a \textit{passage} until its length is at most 100 words. We embed the constituent \textit{passages}, rather than full articles given the context window size limitations of the large language that we use in this work. Furthermore, as argued by Hanley et~al.~\cite{hanley2023specious} and shown by Pikbus et~al.~\cite{piktus2021web}, given that articles often address multiple ideas, embedding \textit{passages} allows us to track the often single idea present within the passage.\footnote{For example, a New York Times article addressing a novel virus spreading in China may also address the Chinese government's previous approaches to dealing with the COVID-19 pandemic.\url{https://web.archive.org/web/20231125001740/https://www.nytimes.com/2023/11/23/world/asia/who-china-respiratory-children.html}} Our dataset consists of 428,051,085~passages.

\section{\label{sec:peft}PEFT through LoRA}
We utilize \textit{Parameter Efficient Fine-Tuning}/PEFT~\cite{lester2021power} through \textit{Low-Rank Adaption}/LoRA~\cite{hu2021lora} to fine-tune and adapt pre-trained models to our datasets or to better their performance. LoRA, specifically, after freezing the weights of the original pre-trained model learns pairs of low-rank-decomposition matrices, reducing the amount of parameters that need to be learned.  LoRA has been shown to often outperform other types of adaptions including full-tuning~\cite{hu2021lora}. Once learned, these matrices are merged with the original frozen weights. LoRA requires the specification of the rank of the matrices learned and an $\alpha$ value that scales the learned parameters. Within this work, we learn LoRA matrices for the attention and the dense/linear layers of our models and utilize the commonly used defaults of rank=8 and $\alpha$=16~\cite{hu2021lora}.

\section{Training\,with\,Unsupervised\,Contrastive\,Loss \label{sec:contrastive}}

To adapt our embedding models to our news dataset, we utilize unsupervised contrastive learning~\cite{gao2021simcse}. For training, this is such that we embed each example $x_i = (passage_i) \in D_{News}$ (where $passage_i$ is the passage text) twice (with dropout both times) with a given model by inputting $[CLS] text_i [SEP]$ and averaging the contextual word vectors of the resulting output as a hidden vector $\mathbf{h}_i$ and $\tilde{\mathbf{h}}_i$  for $passage_i$ as its representations. Then, given a set of hidden vectors $\{\mathbf{h}_i\}_{i=0}^{N_b}$ and $\{\tilde{\mathbf{h}}_{j}\}_{j=0}^{N_b}$ (different dropout), where $N_b$ is the size of the batch, we perform a contrastive learning step for each batch. This is such that for each Batch $\mathcal{B}$, for an \textit{anchor} hidden embedding $\mathbf{h_i}$ within the batch, the set of hidden vectors $\mathbf{h_i} \,, \mathbf{\tilde{h_j}} \in \mathcal{B}$, vectors where $i = j$ are positive pairs. Other pairs where $i\neq j$ are considered negative pairs. Within each batch $\mathcal{B}$, the contrastive loss is computed across all positive pairs in the batch:
\[
    L_{sim} = -\frac{1}{N_b} \sum_{\mathbf{h}_i \in \mathcal{B}}\mathit {l}^c(\mathbf{h}_i )
\]
\[
\mathit{l}^c(\mathbf{h}_i) = \text{log}\frac{ \sum_{j\in\mathcal{B} } \mathbbm{1}_{[i = j]}\mathrm{exp}( \frac{\mathbf{h}_i^\top \tilde{\mathbf{h}_j}}{\tau||\mathbf{h}_i || ||\tilde{\mathbf{h}_j} || })}{\sum_{j\in\mathcal{B}} \, \mathrm{exp}( \frac{\mathbf{h}_i^\top \tilde{\mathbf{h}_j}}{\tau||\mathbf{h}_i || ||\tilde{\mathbf{h}_j} || })}
\]
where, as in prior work~\cite{liang2022jointcl}, we utilize a temperature $\tau=0.07$. When performing fine-tuning, we utilize default hyperparameters (learning rate $3\times 10^{-5}$, batch size=128, and 1M examples) specified in Gao et~al.~\cite{gao2021simcse}.

\section{Pointwise Mutual Information\label{sec:appendix-pmi}}

The PMI of a word $word_i$ in a cluster $C_j$ is calculated:
\begin{align*}
PMI(word_i, C_j) = \text{log}_2\frac{P(word_i,C_j)}{P(word_i) P(c_i)}
\end{align*}

\noindent where $P$ is the probability of occurrence and a scaling parameter $\alpha=1$ is added to the counts of each word per cluster. This scaling parameter $\alpha$ prevents low-frequency words in each cluster from having the highest PMI value~\cite{turney2001mining}. 

\section{Evaluation on SemEval22}

\begin{figure}[h!]
  \centering
  \includegraphics[width=1\columnwidth]{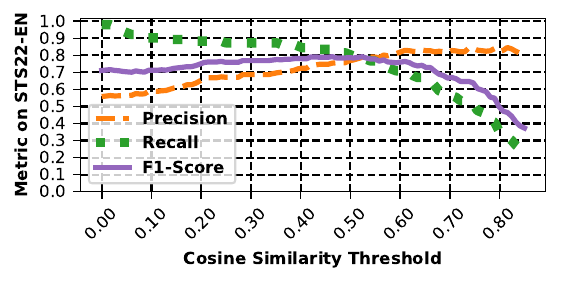}
\caption{Evaluation of our model's precision, recall, and $F_1$ scores on the English portion of the SemEval22 test dataset~\cite{goel2022semeval} (using 3.0 as the cut-off for the two articles being about the same event~\cite{hanley2022partial}).
}
\label{fig:semeval-study}
\end{figure}

\newpage

\section{Passage Pairs at Various Thresholds\label{sec:thresholds}}

\begin{figure}[!h]
\begin{minipage}{.47\textwidth}
\centering \small
\textbf{0.35 Similarity}
\end{minipage}
\noindent\fcolorbox{black}{lightgray}{%
\begin{minipage}{.47\textwidth}
\small
\textbf{PASSAGE 1:} Russias war in Ukraine created a full-on energy crisis as Moscow reduced or cut off natural gas flows to European countries that rely on the fuel to power industry, generate electricity and heat and cool homes. Shrinking supplies, higher demand and fears of a complete Russian cutoff have driven natural gas prices to record highs, further fueling inflation that has squeezed peoples ability to spend and raised the risk of a recession in Europe and the U.K. 

\textbf{PASSAGE 2:}Fingrid had been expecting that Olkiluoto 3 alone would more than compensate for the loss of Russian power imports this winter, he added. Imports of Russian power stopped in May after Russian utility Inter RAO said it had not been paid for the power it sold via pan European exchange Nord Pool since May 6. Without Olkiluoto 3 the situation is quite tight because that would have been more than 10 of the peak demand alone, Ruusunen said. 
\end{minipage}}

\label{figure:all-paragraph_pairs1}
\end{figure}
\begin{figure}[!h]
\begin{minipage}{.47\textwidth}
\centering\small
\textbf{0.40 Similarity}
\end{minipage}
\noindent\fcolorbox{black}{lightgray}{%
\begin{minipage}{.47\textwidth}
\small
\textbf{PASSAGE 1:} HONG KONG, Jan 13 (Reuters Breakingviews) - WM Motor, a Chinese electric-car star that has fallen on hard times, is reversing into an ambitious Plan C. After two earlier attempts at an initial public offering floundered, the group founded by Zhejiang Geely veteran Freeman Shen will at last go public by selling its auto business to Hong Kong-listed Apollo Future Mobility (0860.HK) for \$2 billion. They make an odd couple. Although both are car companies, their cars have little in common. 

\textbf{PASSAGE 2:} Global and Chinese automakers plan to unveil more than a dozen new electric SUVs, sedans and muscle cars this week at the Shanghai auto show, their first full-scale sales event in four years in a market that has become a workshop for developing electrics, self-driving cars and other technology. 
\end{minipage}}

\label{figure:all-paragraph_pairs3}
\end{figure}
\begin{figure}[!h]
\begin{minipage}{.47\textwidth}
\centering\small
\textbf{0.45 Similarity}
\end{minipage}
\noindent\fcolorbox{black}{lightgray}{
\begin{minipage}{.465\textwidth}
\small
\textbf{PASSAGE 1:} Diseases like Ebola and other hemorrhagic fevers were responsible for 70\% of those outbreaks, in addition to illnesses like monkeypox, dengue, anthrax and plague. We must act now to contain zoonotic diseases before they can cause widespread infections and stop Africa from becoming a hotspot for emerging infectious diseases, WHOs Africa director, Dr. Matshidiso Moeti said in a statement.

\textbf{PASSAGE 2:} Equatorial Guinea has confirmed 13 cases of Marburg disease since the beginning of the epidemic, its health officials said on Wednesday after the head of the World Health Organization (WHO) urged the Central African country's government to report new cases officially. Marburg virus disease is a viral haemorrhagic fever that can have a fatality rate of up to 88\%, according to the WHO.
\end{minipage}}

\label{figure:all-paragraph_pairs2}
\end{figure}
\begin{figure}[!h]
\begin{minipage}{.47\textwidth}
\centering\small
\textbf{0.50 Similarity}
\end{minipage}
\noindent\fcolorbox{black}{lightgray}{%
\begin{minipage}{.47\textwidth}
\small
\textbf{PASSAGE 1:} The fatal accident occurred around 4 a.m., and it took several hours to clear the freeway. The firetruck had to be towed away. The Model S was among the nearly 363,000 vehicles Tesla recalled on Thursday because of potential flaws in its Full Self-Driving system. While the recall is aimed at correcting possible problems at intersections and with speed limits, it comes amid a broader investigation by U.S. safety regulators into Teslas automated driving systems.'

\textbf{PASSAGE 2:} The National Highway Traffic Safety Administration (NHTSA) is opening a special investigation into the crash of a 2021 Tesla Model Y vehicle that killed a motorcyclist in California, it said on Monday. Since 2016, NHTSA has opened 37 special investigations of crashes involving Tesla vehicles and where advanced driver assistance systems such as Autopilot were suspected of being used. A total of 18 crash deaths were reported in those Tesla-related investigations, including the most recent fatal California crash. 
\end{minipage}}

\caption{Example of passage pairs at different levels of cosine similarity. }
\label{figure:all-paragraph_pairs4}
\end{figure}

\clearpage

\section{NETINF Algorithm\label{sec:netinf}} 

{NETINF} is a {greedy} algorithm that iteratively computes the marginal gain (\textit{i.e.}, the explanatory power of adding the edge given the set of cascades) of adding a particular weighted edge between two entities/nodes and by only considering the most probable transmission tree (\textit{i.e.}, the steps taken for a given narrative to reach a particular website),  {NETINF} efficiently infers the relationships within the underlying network. The {NETINF} algorithm returns both the marginal gain as well the rate of information flow between two nodes/websites for each edge.  
Thus given a set of cascades $C$, {NETINF} aims to find the the graph $\hat{G}$ that solves the optimization problem~\cite{gomez2012inferring}:
\[
 \hat{G} = \text{arg}\max_{|G| \leq k} P(C|G)
\]
where
\[
P (C|G) = \prod_{c\in C} P(c|G)
\]
\[
P(c|G) = \sum_{T\in T(G)} P(c|T)P(T|G)\propto \sum_{T\in T(G)}\prod_{(i,j)\in T} P_c(i,j)
\]
\noindent
and where $c$ is a cascade, $T(G)$ is the set of all directed spanning trees on $G$, andd $P_c(i,j)$  (or that the node $i$ influences node $j$ in a cascade) is proportional to the time difference between when the two nodes are infected (\textit{i.e.}, when the two websites post a given story about a particular narrative), given an exponential waiting time for infection. 

To optimize this formulation, {NETINF}   only considers the most likely propagation tree $T$ for a given cascade $c$
\[
P (C|G) =\prod_{c\in C}\max_{T\in T{G}}P(c|T) =\prod_{c\in C} \max_{T\in T{G}}\prod_{(i,j)\in T} P_c(i,j)
\]
The improvement of the log-likelihood of a given cascade $c$ for a graph $G$ over the empty graph $K$ is then:
\[
F_c(G) = \max_{T \in T(G)}\text{log} P(c|T) -\max_{T \in T(K)}\text{log} P(c|T)
\]
and the {NETINF} algorithm optimizes the following objective function by iteratively and greedily adding edges with the highest marginal gain to the objective function:
\[
F_C(G) = \sum_{c\in C} F_c(G)
\]

\newpage

\section{Optimized DP-Means}\label{sec:appendix-dpmeans}
DP-Means~\cite{kulis2011revisiting} is a non-parametric extension of the K-means algorithm that does not require the specification of the number of clusters \textit{a priori}. Within DP-Means, when a given datapoint is a chosen parameter $\lambda$ away from the closest cluster, a new cluster is formed. Dinari et al.~\cite{dinari2022revisiting} parallelize this algorithm by \textit{delaying cluster creation} until the end of the assignment step. Namely, instead of creating a new cluster each time a new datapoint is discovered, the algorithm instead determines which datapoint is furthest from the current set of clusters and then creates a new cluster with that datapoint. By delaying cluster creation, the DP-means algorithm can be trivially parallelized. Furthermore, by delaying cluster creation, this version of DP-Means avoids over-clustering the data (\textit{i.e.,} only the most disparate datapoints create new clusters)~\cite{dinari2022revisiting}.
\clearpage
\onecolumn
\section{Evaluation of Clusters\label{sec:cluster-eval}}

\begin{table*}[h]
\begin{minipage}{1.0\textwidth}
\centering
\small
\begin{tabular}{llrr l}
\toprule
 &   &  Passages & \\
Narr. &{Keywords} & Checked & Prec. \\\midrule
1 & laissez-faire, progressivism, liberalism, laissez, corpus&193 &96.89\% \\
2 & quake, earthquake, aftershock, turkey, rubble& 500 & 100.00\% \\
3 & sinema, manchin, filibuster, kyrsten, senate& 500 & 100.00\%\\
4 & williamson, marianne, self-help, williamsons, sander& 500 & 97.40\% \\
5 &dysphoria, puberty, blocker, crosssex, hormone & 500 & 100.00\%\\
6 &sudan, anand, evacuation, sudanese, khartoum & 500 & 100.00\%\\
7 &rioter, slogan, bearing, capitol, drum & 500 & 99.20\%\\
8 &teixeira, dighton, guardsman, teixeiras, massachusetts & 500 & 99.40\%\\
9 &fdny, firefighter, firehouse, klein, kavanagh & 500 & 97.00\% \\
10 &bragg, alvin, rouser, nypd, rabble & 500 & 95.60\%\\
11&taliban, afghan, afghanistan, hunger, malnutritio & 500 &100.00\%\\
12&eyesight, blindness, blind, eye, sight& 500 & 99.40\%\\
13 & maralago, classified, ballroom, fundraiser, document & 500 & 100.0\% \\
14&carolina, vetoproof, map, raleigh, cooper & 500 & 99.80\% \\
15& tarantino, quentin, pulp, cinema, filmmaker & 500 & 99.20\%\\
16 &seoul,korea, posco, compensate, keb &500 & 100.00\% \\
17& miscarriage, pregnant, pregnancy, csection, motherhood& 500 &  100.00\% \\
18& faucis, niaid, anthony, gain-of-function, allergy& 500 & 100.00\% \\
19& crump, arbery, breonna, ahmaud, trayvon& 500 & 99.08\%\\
20&portuguese, slave, plantation, colony, dutch& 500 & 100.00\%\\
21& cadet, guard, harassment, assault, adjutant & 500 & 100.00\% \\
22 & spam, bot, musk, twitter, elon& 500 & 98.80\%\\
23 &ufo, roswell, sighting, saucer, alien& 500 & 100.00\%\\
24 & cpu, intel, x86, processor, amd& 500 &96.40\%\\
25& chappelle, comedian, isaiah, onstage, attacker& 500& 100.00\%\\
26& burisma, pozharskyi, vadym, hunter, zlochevsky& 500 & 100.00\%\\
27&ubridgerton, penelope, featherington, daphne, coughland&  500 & 100.00\% \\
28&naloxone, narcan, over-the-counter, emergent, nasal & 500 &100.00\%  \\
29&schmitt, greitens, hartzler, missouri, trudy& 100 & 99.20\%\\
30&currency, dollar, yuan, reserve, de-dollarization& 500 &99.80\% \\
\bottomrule
 && \textbf{Prec.}& \textbf{99.26\%}
\end{tabular}
\end{minipage}
\end{table*}


\clearpage
\newpage
\section{Additional Stances\label{sec:additional-stances}}

\begin{table}[htbp]
\centering
\small
\begin{tabular}{l|c|c}
{Pro-China Stances} & {Coeff.} & {Std.} \\\midrule
Pro self-reliance & 0.286 & 0.092 \\ 
Pro communist & 0.180 & 0.093 \\ 
Pro africa & 0.126 & 0.103 \\ 
\midrule
{Anti-China Stances}  \\\midrule
Against communist & -0.288 & 0.087 \\
Against covid & -0.228 & 0.091 \\
Against russia & -0.215 & 0.072 \\
\hline
\end{tabular}
\caption{The stances associated with China according to the Bayesian model.}
\end{table}

\begin{table}[htbp]
\centering
\small
\begin{tabular}{l|c|c}
{Pro-Iran Stances} & {Coeff.} & {Std.} \\\midrule
Against yemeni & 0.055 & 0.061 \\ 
Pro islamic & 0.052 & 0.042 \\ 
Pro armenia & 0.049 & 0.036 \\  
\midrule
{Anti-Iran Stances}  \\\midrule
Against tehran & -0.073 & 0.060 \\
Against islamic & -0.058 & 0.043 \\
Against communist & -0.047 & 0.038 \\
\hline
\end{tabular}
\caption{The stances associated with Iran according to the Bayesian model.}
\end{table}

\begin{table}[htbp]
\centering
\small
\begin{tabular}{l|c|c}
{Pro-America Stances} & {Coeff.} & {Std.} \\\midrule
Pro harvard & 0.111 & 0.064 \\ 
Pro allstar & 0.085 & 0.055 \\ 
Pro mvp & 0.070 & 0.045 \\ 
\midrule
{Anti-America Stances}  \\\midrule
Against graham & -0.143 & 0.070 \\
Against cia & -0.126 & 0.069 \\
Against canada & -0.126 & 0.060 \\
\hline
\end{tabular}
\caption{The stances associated with America according to the Bayesian model.}
\end{table}

\begin{table*}[!ht]
\begin{minipage}{1.0\textwidth}
\centering
\small
\begin{tabular}{lll}
\multicolumn{3}{c}{\textbf{ Pro America}} \\
\toprule
Reliable & Mixed & Unreliable \\ \midrule
messengernews.net (100.0\%) & bluegrasstimes.com (100.0\%)  & adflegal.org (100.0\%)  \\ 
nationalgeographic.com (100.0\%) & tribunnews.com (100.0\%)  & worldcouncilforhealth.org (100.0\%)  \\ 
fee.org (100.0\%) & yaf.org (100.0\%)  & nvic.org (100.0\%)  \\    \\ 
\bottomrule
\multicolumn{3}{c}{\textbf{ Against America}} \\
\toprule
Reliable & Mixed & Unreliable \\ \midrule
thepeoplescube.com (100.0\%) & thepatriotjournal.com (76.9\%)  & infostormer.com (80.0\%)  \\ 
cjr.org (50.0\%) & greanvillepost.com (68.9\%)  & paulcraigroberts.org (71.4\%)  \\ 
techdirt.com (46.2\%) & newscorpse.com (58.8\%)  & gellerreport.com (64.9\%)  \\ 
\bottomrule
\end{tabular}
\end{minipage}

\begin{minipage}{1.0\textwidth}
\centering
\small

\begin{tabular}{lll}
\multicolumn{3}{c}{\textbf{Pro China}} \\
\toprule
Reliable & Mixed & Unreliable \\ \midrule
dailyhive.com (64.1\%) & cctv.com (55.1\%)  & discoverthenetworks.org (50.0\%)  \\ 
emerging-europe.com (64.0\%) & egypttoday.com (47.1\%)  & occupydemocrats.com (41.7\%)  \\ 
tdn.com (58.3\%) & thecountersignal.com (45.5\%)  & medicalkidnap.com (37.8\%)  \\ 
\bottomrule

\multicolumn{3}{c}{\textbf{ Against China}} \\
\toprule
Reliable & Mixed & Unreliable \\ \midrule
tabletmag.com (70.0\%) & boundingintocomics.com (100.0\%)  & voterig.com (94.6\%)  \\ 
outsidethebeltway.com (69.8\%) & faithwire.com (81.8\%)  & govtslaves.com (93.3\%)  \\ 
icij.org (63.6\%) & fff.org (81.8\%)  & conservativeplaylist.com (82.4\%)  \\ 
\bottomrule
\end{tabular}
\end{minipage}

\begin{minipage}{1.0\textwidth}
\centering
\small
\begin{tabular}{lll}
\multicolumn{3}{c}{\textbf{Pro Iran}} \\
\toprule
Reliable & Mixed & Unreliable \\ \midrule
cambridge.org (55.9\%) & tasnimnews.com (36.9\%)  & presstv.ir (34.8\%)  \\ 
lamag.com (50.0\%) & almanar.com.lb (34.5\%)  & theconservativetreehouse.com (33.3\%)  \\ 
msmagazine.com (46.4\%) & mehrnews.com (32.6\%)  & healthimpactnews.com (25.0\%)  \\ 
\bottomrule

\multicolumn{3}{c}{\textbf{Against Iran}} \\
\toprule
Reliable & Mixed & Unreliable \\ \midrule
castanet.net (78.1\%) & thefederalist.com (78.6\%)  & hagmannreport.com (100.0\%)  \\ 
dailysignal.com (72.2\%) & smirkingchimp.com (78.6\%)  & clarionproject.org (69.2\%)  \\ 
wearethemighty.com (70.0\%) & patriotnewsalerts.com (76.9\%)  & thewashingtonstandard.com (68.8\%)  \\ 
\bottomrule
\end{tabular}
\end{minipage}

\caption{The set of news websites with the highest percentage of  \textit{Pro}-articles about various entities/topics.}
\end{table*}

\begin{figure*}[!htbp]

  \begin{subfigure}[b]{0.5\textwidth}
        \centering
        \includegraphics[width=\textwidth]{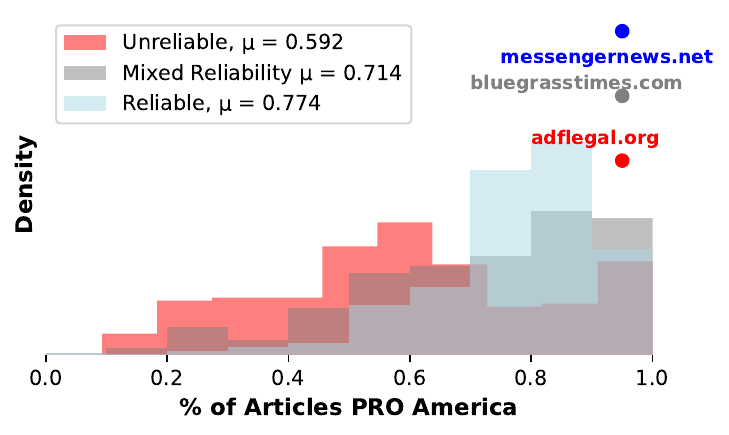}
        \caption{Dist. of Pro America}
        \label{fig:d4-pro-america}
    \end{subfigure}
      \begin{subfigure}[b]{0.5\textwidth}
        \centering
        \includegraphics[width=\textwidth]{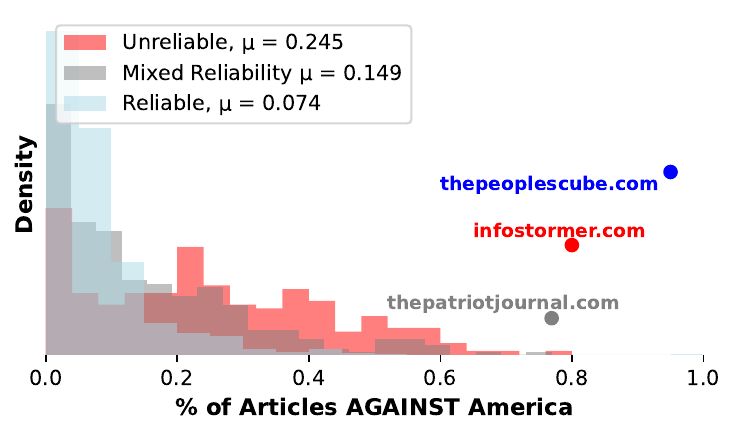}
        \caption{Dist. of Against America}
        \label{fig:d4-against-america}
    \end{subfigure}
    
    \begin{subfigure}[b]{0.5\textwidth}
        \centering
        \includegraphics[width=\textwidth]{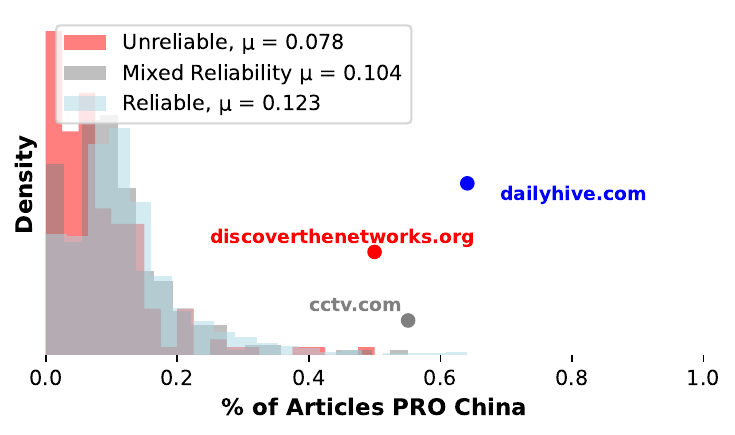}
        \caption{Dist. of Pro China}
        \label{fig:d4-pro-china}
    \end{subfigure}
    \begin{subfigure}[b]{0.5\textwidth}
        \centering
        \includegraphics[width=\textwidth]{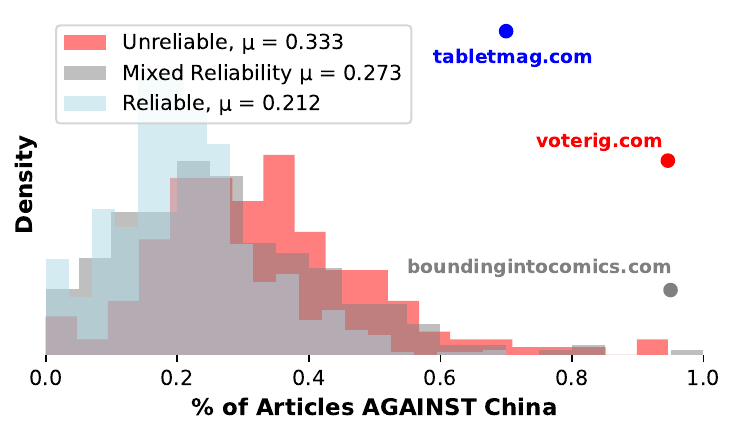}
        \caption{Dist. of Against China}
        \label{fig:d4-against-china}
    \end{subfigure}
    
    \begin{subfigure}[b]{0.5\textwidth}
        \centering
        \includegraphics[width=\textwidth]{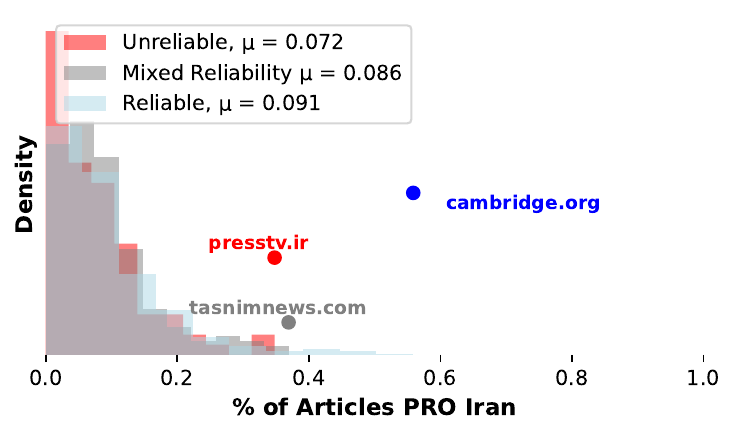}
        \caption{Dist. of Pro Iran}
        \label{fig:d4-pro-iran}
    \end{subfigure}
    \begin{subfigure}[b]{0.5\textwidth}
        \centering
        \includegraphics[width=\textwidth]{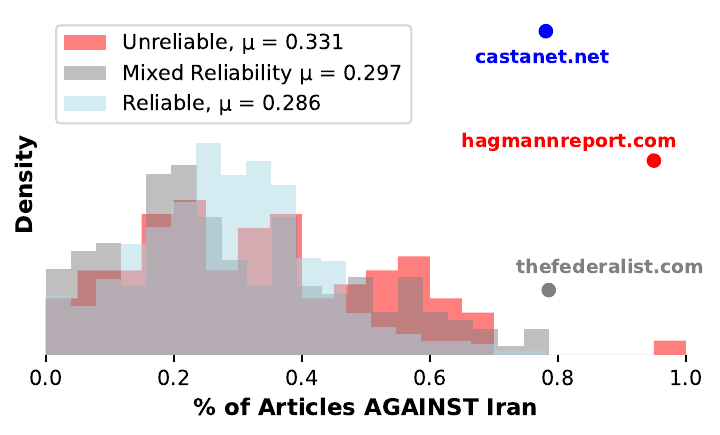}
        \caption{Dist. of Against Iran}
        \label{fig:d4-iran}
    \end{subfigure}

 \centering

    \caption{Distribution of stances to various entities.}
    \label{fig:various-entities-other}
\end{figure*}

\begin{figure*}[!htbp]
 \begin{subfigure}[b]{0.5\textwidth}
        \centering
        \includegraphics[width=\textwidth]{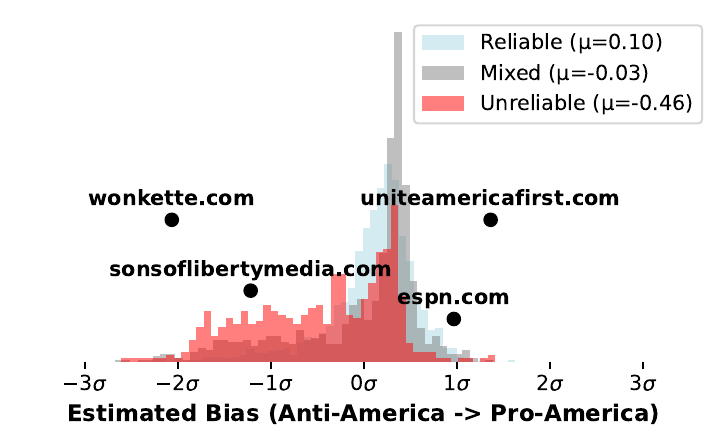}
        \caption{America Latent}
        \label{fig:america-latent}
    \end{subfigure}
    \begin{subfigure}[b]{0.5\textwidth}
        \centering
        \includegraphics[width=\textwidth]{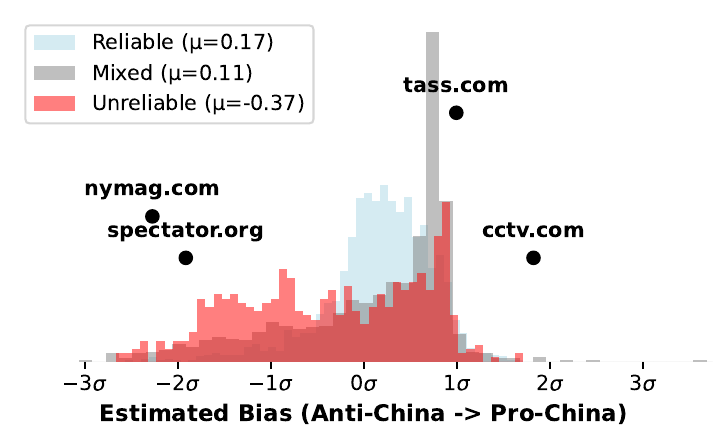}
        \caption{China Latent}
        \label{fig:china-latent}
    \end{subfigure}
    \centering
    \begin{subfigure}[b]{0.5\textwidth}
        \centering
        \includegraphics[width=\textwidth]{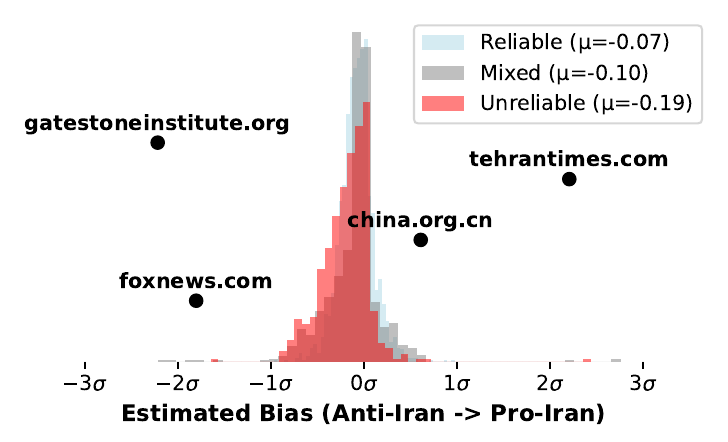}
        \caption{Iran Latent}
        \label{fig:iran-latent}
    \end{subfigure}

 \centering

    \caption{Latents for Biases to entities.}
    \label{fig:additional-latents}
\end{figure*}

\newpage
\clearpage

\onecolumn
\section{Articles over Time\label{sec:articles-over-time}}

\begin{table*}
\centering
\small
\begin{tabularx}{1.0\columnwidth}{srrX}
\toprule
\multicolumn{4}{c}{\textbf{Reliable News}} \\ 
\midrule
{Keywords} & Articles & Websites & Most Prolific Domains\\ 
\midrule
desantis, ron, florida, gov, governor  & 62,911 & 1,330 & yahoo.com (5,454), floridapolitics.com (3,098), thehill.com (1,900)\\ 
quarter, net, revenue, expense, loss & 57,453 & 824 & benzinga.com (19,266), yahoo.com (3,356), marketwatch.com (3,106)\\ 
half, halftime, minute, goal, possession & 52,362 & 809 & yahoo.com (2,785), espn.com (2,527), rte.ie (2,124)\\ 
symptom, fever, cough, throat, sore & 42,581 & 1,241 & yahoo.com (1,685), webmd.com (1,023), news-medical.net (869)\\ 
roe, wade, abortion, dobbs, supreme & 38,358 & 1,317 & news-yahoo.com (2,118), thehill.com (798), cnn.com (615)\\ 
\midrule
\multicolumn{4}{c}{\textbf{Mixed-Reliability News}} \\  \midrule
{Keywords} & Articles & Websites & Most Prolific Domains\\ 

\midrule
desantis, ron, florida, gov, governor & 32,723 & 386 & tampafp.com (1,762), theepochtimes.com (1,441), nypost.com (1,220)\\ 
half, halftime, minute, goal, possession & 21,113 & 228 & the-sun.com (2,402), thesun.co.uk (2,354), independent.co.uk (1,968)\\ 
kourtney, kardashian, travis, khloe, barker & 19,226 & 117 & the-sun.com (7,046), thesun.co.uk (4,951), metro.co.uk (1,028)\\ 
ronaldo, cristiano, portugal, football, manchester & 17,791 & 126 & the-sun.com (3,307), thesun.co.uk (3,024), dailystar.co.uk (1,732)\\ 
kardashian, khloe, kim, kourtney, jenner & 16,578 & 145 & the-sun.com (5,603), thesun.co.uk (3,781), metro.co.uk (987)\\ 
\midrule
\multicolumn{4}{c}{\textbf{Unreliable News}} \\ \midrule
{Keywords} & Articles & Websites & Most Prolific Domains\\ 

\midrule
desantis, ron, florida, gov, governor  & 13,963 & 328 & dailymail.co.uk (2,831), ussanews.com (902), occupydemocrats.com (496)\\ 
dictatorship, democracy, censorship, propaganda, authoritarian & 9,094 & 358 & abovetopsecret.com (2,549), noqreport.com (384), theburningplatform.com (293)\\ 
respect, dignity, equal, equality, freedom & 7,950 & 333 & dailymail.co.uk (3,090), abovetopsecret.com (2,183), lifesitenews.com (160)\\ 
classified, document, maralago, archive, wilmington & 7,678 & 249 & dailymail.co.uk (1,393), thegatewaypundit.com (658), survivethenews.com (253)\\ 
constitution, oath, amendment, constitutional, defend & 7,655 & 353 & abovetopsecret.com (2,044), dailymail.co.uk (452), theqtree.com (307)\\ 
\bottomrule
\end{tabularx}
\caption{\label{tab:narratives_combined} Top Narratives from Different News Sources by Number of Articles in Our Dataset.}
\end{table*}

\begin{figure*}[h]
 \centering
   
    \begin{subfigure}[b]{0.90\textwidth}
        \centering
        \includegraphics[width=\textwidth]{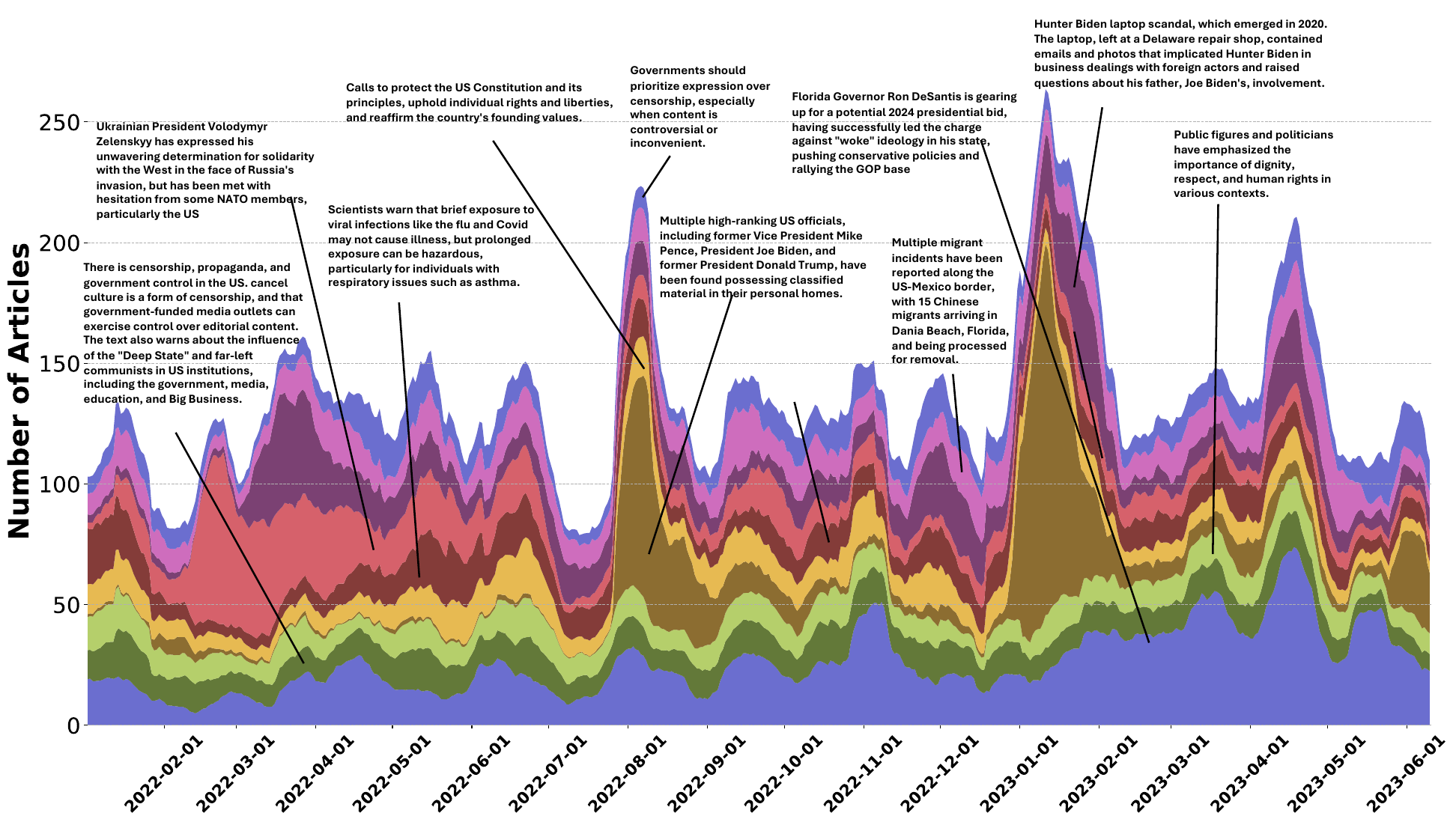}
        \caption{Story Spread on Unreliable News Websites}
    \end{subfigure}

        \label{fig:unreliable-topics}
\end{figure*}

\begin{figure*}[!htbp]
 \centering
    \begin{subfigure}[b]{0.9\textwidth}
        \centering
        \includegraphics[width=\textwidth]{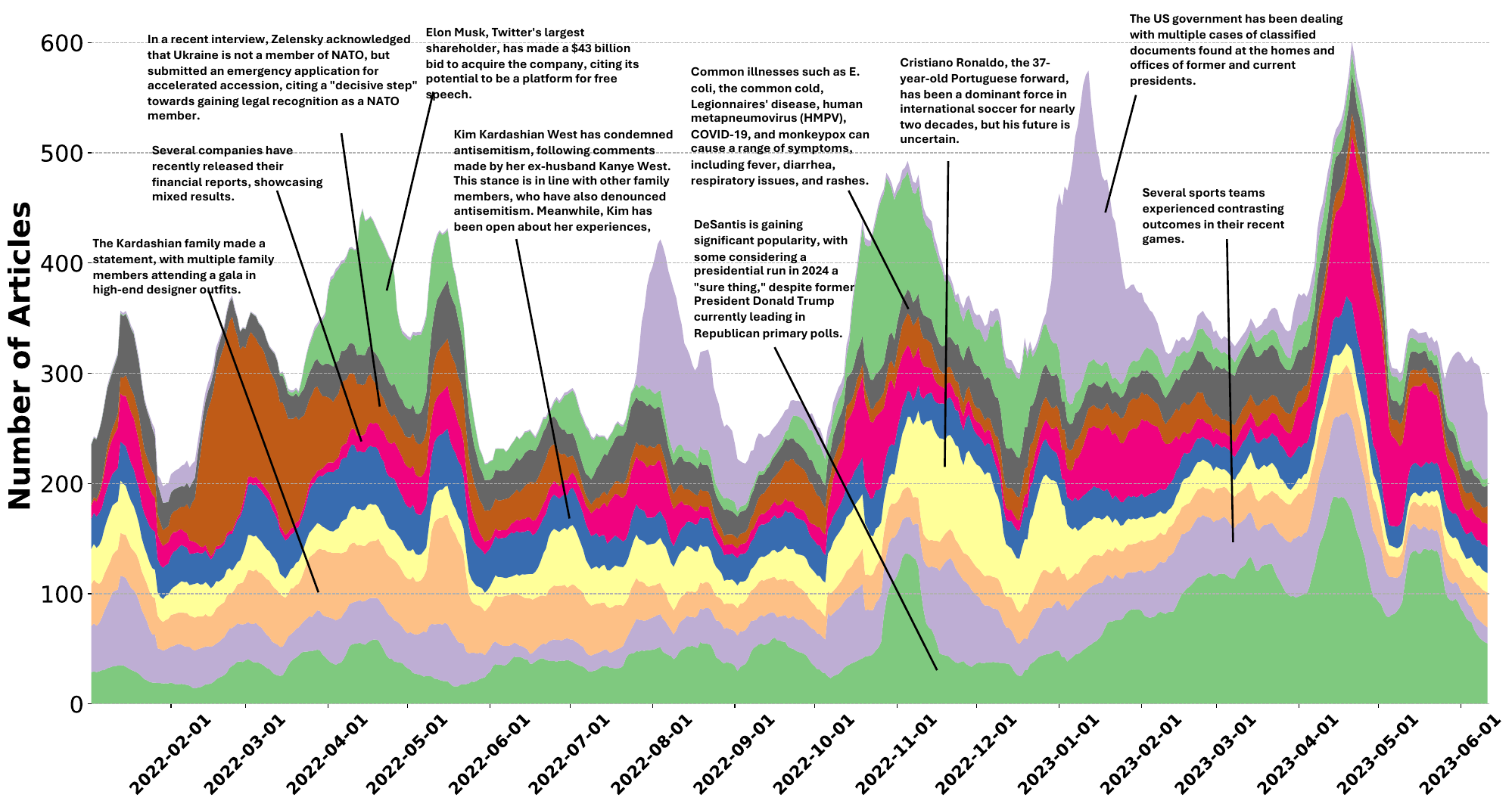}
      \caption{Story Spread on Mixed-Reliability News Websites}
    \end{subfigure}
        \label{fig:mixed-topics}
\end{figure*}

\clearpage

\onecolumn
\section{Keywords and Auto-Generated Summaries\label{sec:cluster-spec-additional-summaries}}

\begin{table*}[h!]
\centering

\small
\begin{tabularx}{1\textwidth}{lsX}
Story. &{Keywords} &\\\midrule
& & Auto-Generated Summary   \\\cline{3-3} 
1 & transgender, trans, minor, banning, restrict& Across the United States, anti-trans legislation is on the rise, with multiple states imposing restrictions on transgender individuals, including bans on gender-affirming care for minors, drag shows, and participation in female school sports. At least 20 states have imposed similar limits on trans athletes, and nearly half of all states have laws that make it difficult for transgender youth to receive necessary care. Companies and lawmakers are being urged to take a stance against these laws, which are seen as a personal attack on the LGBTQ+ community.  \\ \cline{3-3} 

& & Random Sample Passage  \\\cline{3-3} 

 & & Contino described how anti-trans legislation has grown across the country, including two bills in Tennessee that ban gender-affirming care for minors as well as drag shows. She says the overarching impact this culture can have on transgender people is huge, and companies need to take a stance.  \\ \hline 
& & Auto-Generated Summary   \\\cline{3-3} 
 2 & buckingham, mall, palace, procession, royal& Large crowds have gathered in London to pay respects to the Queen, with many camping out for days to secure a prime viewing spot for the procession and potentially catch a glimpse of the newly crowned monarch and Queen Camilla.  \\ \cline{3-3} 

& & Random Sample Passage  \\\cline{3-3} 

 & & Some 100 heads of state, representatives from 200 countries and hundreds of thousands of visitors are expected to descend on London for the historic event. Many die-hard royal fans are already camped out near Buckingham Palace to secure the best viewing spot.  \\ \hline 

 & & Auto-Generated Summary   \\\cline{3-3} 
 3 & card, debit, atm, credit, fraudulent& To protect against credit card theft and unauthorized transactions, it's essential to keep important documents secure and be aware of potential scams, such as card skimming. Credit card data can also indicate changes in consumer spending habits. However, the use of credit and debit cards has become increasingly popular, with many retailers now accepting these forms of payment. Despite the convenience of these cards, there is still a risk of theft and unauthorized transactions, with some cases involving large-scale data breaches and individuals using sleight-of-hand techniques to steal gift cards.  \\ \cline{3-3} 

& & Random Sample Passage  \\\cline{3-3} 

 & & Card skimming involves the use of a device that looks like a normal part of a point-of-sale machine or PIN pad but instead copies EBT card information for the thief. These card skimmer devices are difficult to spot, to the point that retailers may not be aware of their installation.\\ 







\bottomrule
\end{tabularx}
\end{table*}

\
\end{document}